\documentclass[fleqn,usenatbib]{mnras}

\usepackage{newtxtext,newtxmath}

\usepackage{todonotes}

\usepackage[T1]{fontenc}

\DeclareRobustCommand{\VAN}[3]{#2}
\let\VANthebibliography\thebibliography
\def\thebibliography{\DeclareRobustCommand{\VAN}[3]{##3}\VANthebibliography}

\usepackage{graphicx}	
\usepackage{amsmath}	
\defcitealias{mat15CVdiskwind}{M15} 
\defcitealias{lon02python}{LK02} 
\defcitealias{tam22v455andspec}{T22} 
\defcitealias{ara05v455and}{A05}

\def\python{\textsc{Python}}
\def\heiifses{He~\textsc{ii}~$\lambda4686$}
\def\heiiosfo{He~\textsc{ii}~$\lambda1640$}
\def\heiittot{He~\textsc{ii}~$\lambda3202$}
\def\la{Ly$\alpha$}
\def\ha{H$\alpha$}
\def\hb{H$\beta$}
\def\civoffo{C~\textsc{iv}~$\lambda1550$}
\def\ovotso{O~\textsc{v}~$\lambda$1371}

\title[Disc winds in dwarf novae]{A disc wind origin for the optical spectra of dwarf novae in outburst}

\author[Y. Tampo et al.]{
Yusuke Tampo,$^{1,2}$\thanks{E-mail: tampo@kusastro.kyoto-u.ac.jp (YT)}
Christian Knigge,$^{2}$
Knox S. Long,$^{3,4}$
James H. Matthews,$^{5}$
and Noel Castro Segura$^{2}$
\\
$^{1}$Department of Astronomy, Kyoto University, Kitashirakawa-Oiwake-cho, Sakyo-ku, Kyoto, 606-8502, Japan\\
$^{2}$School of Physics \& Astronomy, University of Southampton, Southampton SO17 1BJ, UK\\
$^{3}$Space Telescope Science Insitute, 3700 San Martin Drive, Baltimore, Maryland 20218, USA\\
$^{4}$Eureka Scientific, Inc. 2452 Delmer Street, Suite 100, Oakland, CA 94602-3017, USA \\
$^{5}$Department of Physics, Astrophysics, University of Oxford, Denys Wilkinson Building, Keble Road, Oxford, OX1 3RH, UK
}

\date{\today}

\pubyear{2024}

\begin{document}
\label{firstpage}
\pagerange{\pageref{firstpage}--\pageref{lastpage}}
\maketitle

\begin{abstract}
Many high-state cataclysmic variables (CVs) exhibit blue-shifted absorption features in their ultraviolet (UV) spectra -- a smoking-gun signature of outflows. 
However, the impact of these outflows on {\em optical} spectra remains much more uncertain. 
During its recent outburst, the eclipsing dwarf nova V455 And displayed strong optical emission lines whose cores were narrower than expected from a Keplerian disc. 
Here, we explore whether disc + wind models developed for matching UV observations of CVs can also account for these optical spectra. 
Importantly, V455~And was extremely bright at outburst maximum: the accretion rate implied by fitting the optical continuum with a standard disc model is $\dot{M}_{\rm acc} \simeq  10^{-7}~{\rm M}_\odot~{\rm yr^{-1}}$. 
Allowing for continuum reprocessing in the outflow helps to relax this constraint. 
A disk wind can also broadly reproduce the optical emission lines, but only if the wind is (i) highly mass-loaded, with a mass-loss rate reaching $\dot{M}_{\rm wind} \simeq 0.4 \dot{M}_{\rm acc}$, and/or (ii) clumpy, with a volume filling factor $f_V \simeq 0.1$. 
The same models can describe the spectral evolution across the outburst, simply by lowering $\dot{M}_{\rm acc}$ and $\dot{M}_{\rm wind}$. 
Extending these models to lower inclinations and into the UV produces spectra consistent with those observed in face-on high-state CVs. 
We also find, for the first time in simulations of this type, P-Cygni-like absorption features in the Balmer series, as have been observed in both CVs and X-ray binaries. 
Overall, dense disc winds provide a promising framework for explaining multiple observational signatures seen in high-state CVs, but theoretical challenges persist.

\end{abstract}




\section{Introduction}
\label{sec:1}

Cataclysmic variables (CVs) are close binary systems containing a primary accreting white dwarf (WD) and a secondary star that fills its Roche lobe (see, e.g. \citealt{war95book, hel01book} for a review).
If the WD has a relatively weak magnetic field, the gas transferred from the secondary forms an accretion disc around the WD.
Dwarf novae (DNe) in outburst and nova-like (NL) variables,
collectively referred to as high-state CVs, are subtypes of CVs whose ultraviolet (UV) -- optical continuum is reasonably approximated by the steady standard disc model \citep{sha73alphadisk} with an accretion rate $\dot{M}_{\rm acc} \sim 10^{-8}$ M$_\odot$ yr$^{-1}$.
\footnote{While some literatures, including recently \cite{god17NLdisk},  have argued for departures from a standard disk mode in disc-dominated high state CVs, many, if not all,  of these discrepancies can be attributed to uncertainties in the atmospheres of the disc, see \citealt{hub21ixvel}.} 
NLs normally stay in a bright state with a high disc accretion rate, while DNe show recurrent outbursts reflecting a sudden increase of the accretion rate due to disc instability (see \citealt{osa96review, las01DIDNXT, kim20thesis, ham20CVreview} for reviews).
In terms of the basic physical characteristics of the accretion disc, such as the accretion rate ($\sim 10^{-8}$ M$_\odot$ yr$^{-1}$), temperature ($\sim 10^{4-5}$ K) and radius (a few tens $\times$ R$_{\rm WD}$ where R$_{\rm WD}$ is the WD radius), the discs in NLs and DN outbursts are thought to be identical to each other.

As first shown in observations obtained with the  International Ultraviolet Explorer, UV spectra of high-state CVs contain P-Cygni-like profiles of \civoffo~ and other resonance lines (e.g., \citealt{hea78iuespec, kra81CVIUE, cor82outburstwind, lad91CVIUE, lon94ixvel, pri95CVwind, kni97zcam}), 
which is bona-fide evidence for the presence of winds (see \citealt{mau97cvwindreview, fro05cvoutflow, pro05cvoutflow} for reviews of CV winds).
By analysing and modelling these UV resonance lines, the geometries and kinematics of biconical outflows arising from standard Shakura-Sunyaev disks have been studied.
For example, high-inclination systems show narrower line profiles during eclipses than away from eclipses.
\citet{shl96v347pup, kni97uxumawindmapping} (and references therein) interpret this as the occultation of the inner rapidly-rotating disc wind, which rotates faster than the outer disc wind considering conservation of angular momentum.
The P-Cygni profiles of CV winds are slightly different from those of stellar winds; the deepest absorption is located at $\leq 2000$ km s$^{-1}$ from the rest wavelength rather than around the bluest edge.
This point suggests a slow acceleration in CV disc winds (e.g., \citealt{dre87CVlineprofile, mau87hlcmaIUE}).  There is no direct observational method to measure the mass-loss rate, but 
through modelling the UV spectrum, various authors \citep[][hereafter \citetalias{lon02python}]{shl93CVwind, vit93cvwind, kni95cvwind, lon02python} have found that the typical ratio of the wind mass-loss rate to the disc accretion rate is 0.01-0.1.
The wind acceleration mechanism is thought to be line-driving, although hydrodynamic simulations have failed to achieve the required mass-loss rates \citep{per97linewind, pro98radiationwind, pro99radiationwind, hig24linedriven}.
These calculations assume a steady disc with $\dot{M}_{\rm acc} = 10^{-9}$ -- $10^{-7}$ M$_\odot$ yr$^{-1}$ and typically yield a ratio of the mass-loss rate to the accretion rate of $10^{-5}$ -- $10^{-3}$.
We note that, more recently, P-Cygni profiles have been detected in some NLs in optical (e.g., \citealt{kaf04windfromCV, cun23nloutflow}) and perhaps in X-ray spectra \citep{bal22bzcamv592casxray} as well.
However, optical and X-ray P-Cygni profiles have never been observed in DN outbursts to the best of our knowledge.

Disc winds have also been invoked to explain the puzzling behaviour of the so-called SW Sex-type NLs.
These are defined as eclipsing NLs that exhibit strong single-peaked Balmer and \heiifses~ emission lines in their optical spectra (e.g., \citealt{tho91pxand, tho91bhlyn, mar99lspeg, sch15swsex}).  
Since line formation in a Keplerian disc viewed at high inclinations cannot explain these single-peaked profiles, disc winds offer a promising alternative interpretation of these peculiar line profiles (\citealt{hon86swsex, hoa94CVHeII, hel96v1315aql, mur96diskwind, mur97diskwind, rib08CVflickering, mat15CVdiskwind}; hereafter \citetalias{mat15CVdiskwind}). 
We note that other mechanisms, such as an accretion curtain of the magnetic WD and/or an elongated hotspot, have also been suggested \citep{wil89CVeclipse, hoa03dwuma, dhi13swsexenigma}.

Recently, \citet[][hereafter \citetalias{tam22v455andspec}]{tam22v455andspec}  presented a series of optical spectra of an eclipsing DN V455~And obtained during its 2007 superoutburst. 
These exhibited strong, narrow, and (in some cases) single-peaked emission lines at the outburst maximum, similar to those in SW Sex-type NLs. 
These line profiles are hard to understand if the lines are formed purely in the atmosphere of a standard geometrically-thin accretion disc. It is therefore natural to ask if line formation in a disc wind may help to explain these observations.

In this paper, we present the state-of-the-art Monte Carlo ionization and radiative transfer simulations of disc winds associated with a DN outburst. 
Our goal is to determine the degree to which disc wind models similar to those previously used to reproduce the UV spectra of high-state CVs can reproduce the optical spectra of V455~And around its outburst maximum. 

The remainder of this paper is organized as follows:  
In  Section \ref{sec:2}, we first summarize what is known about V445 And and then describe the spectral features that \citetalias{tam22v455andspec} ascribed to a disc wind during its superoutburst. 
In Section \ref{sec:3}, we provide a brief description of the radiative transfer code \python\ as used in this effort. 
In Section \ref{sec:4}, we present the results of our simulations.
We discuss the nature of the disc wind in V455~And and its implication for other high-state CVs -- and for the line-driving mechanism more generally -- in Section \ref{sec:5}.
Finally, in Section \ref{sec:summary}, we provide a summary of our main results.

\section{The observational data set and the peak accretion rate}
\label{sec:2}

\begin{figure}
    \centering
    \includegraphics[width=\linewidth]{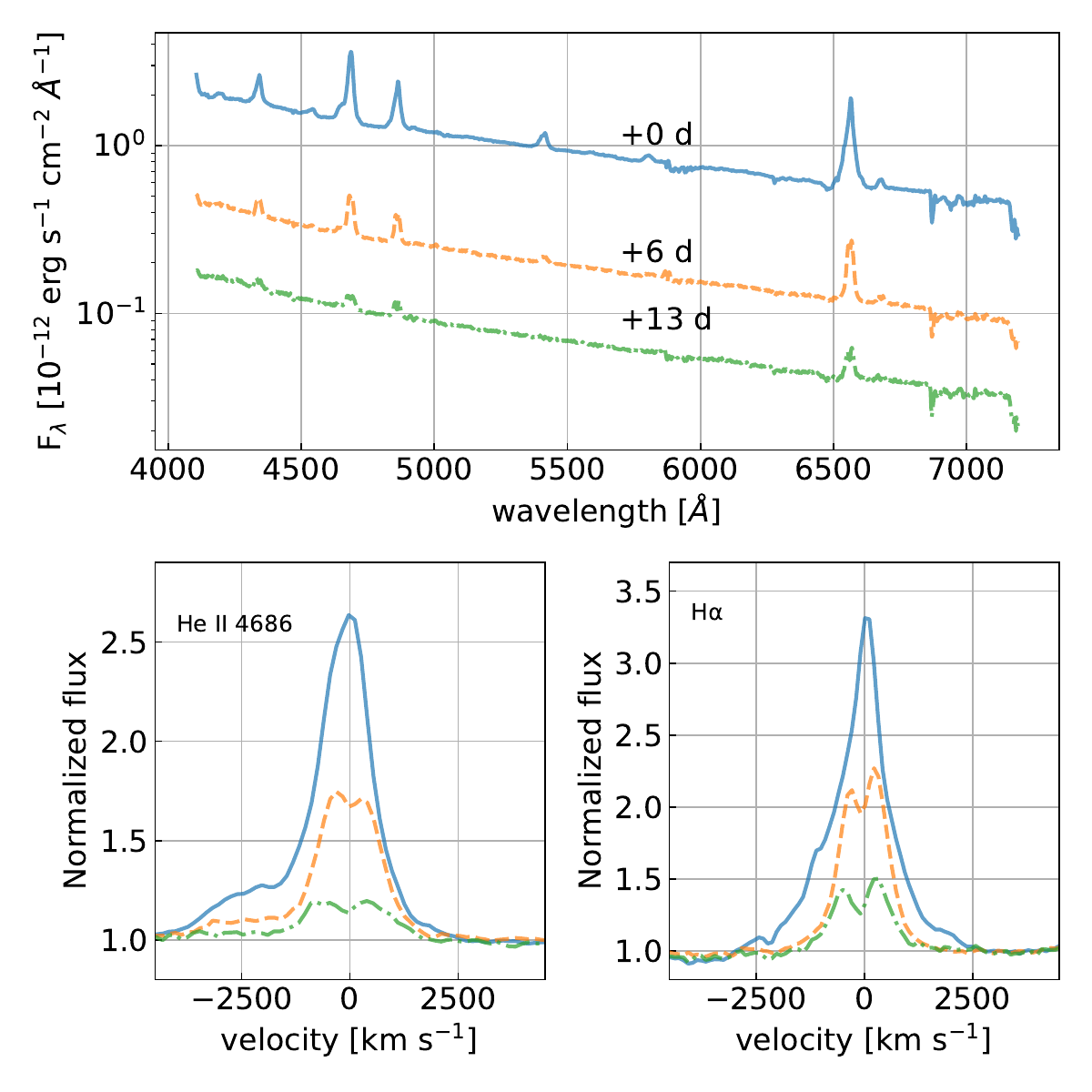}
    \caption{
    The optical spectra of V455~And during its 2007 superoutburst.
    The top panel shows the flux-calibrated spectra on BJD 2454350.04, 2454356.08, and 2454363.03 from upper to lower.
    The bottom panels represent the normalized and zoomed spectra around \heiifses~ (left) and H$\alpha$ (right).
    The original data was taken from \citetalias{tam22v455andspec}.
    }
    \label{fig:v455andspec}
\end{figure}

V455~And is a short-orbital-period (P$_{\rm orb} = 0.05630921(1)$ d; \citealt{ara05v455and}; hereafter \citetalias{ara05v455and}) and low-mass-ratio CV ($q \leq 0.1$; \citetalias{ara05v455and}; \citealt{kat13qfromstageA}). 
The WD temperature of V455 And in quiescence is consistent with that expected from the time-averaged accretion rate of $3 \times 10^{-11}$ M$_\odot$ yr$^{-1}$ \citep{szk13V455And}.
It underwent the WZ Sge-type DN outburst in 2007 and reached $\approx$ 8.6 mag in optical at its outburst maximum  \citep{Pdot, kat15wzsge}.
Time-resolved photometry has revealed grazing eclipses both in quiescence \citepalias{ara05v455and} and during the outburst \citep{Pdot, mat09v455and}, indicating an inclination $i \approx 75^\circ$ \citepalias{ara05v455and}.
According to the Gaia Early Data Release 3, the distance of V455~And is 75.6$^{+0.3}_{-0.2}$ pc \citep{gaiaedr3, Bai21GaiaEDR3distance}.
Assuming this distance and the binary parameters of V455~And (see also table \ref{tab:binaryparam}), the absolute magnitude at optical maximum, M$_V \simeq$ 4.2 mag, implies a very high disc accretion rate of $\sim$ 10$^{-7}$ M$_\odot$ yr$^{-1}$ according to the standard disc model. 

This accretion rate estimate is interesting, because it lies in the region where steady hydrogen burning might take place during outbursts \cite[c.f.][]{kin03dntypeiasn}. 
However, whether nuclear burning would {\em actually} be triggered presumably depends on factors such as the mass and temperature of the hydrogen-rich envelope on the WD (and hence on the time since the last nova eruption of the source). 
These issues are beyond the scope of our investigation here, although we will explore some wind models in which a luminous, soft X-ray emitting boundary layer (BL) is included, with $L_{\rm BL} = L_{\rm disc}$ (i.e., soft X-ray BL model in high-state CVs; \citealt{pri77softXinDN}). That said, if steady burning is, in fact, taking place in V455~And during the outburst, the expected soft X-ray and bolometric luminosities could be around two orders of magnitude higher than even this. A typical optical absolute magnitude of supersoft X-ray sources in \citet{sim03ssxsoptical} is indeed $\geq$ 2 mag brighter than that of V455 And as well. 
Unfortunately, the only available UV or soft X-ray observations were during the outburst decline (\citealt{sen08v455andhardX}; see section \ref{sec:343}), which makes it more difficult to constrain the peak accretion rate and X-ray luminosity.

As we shall see, our own modelling of the continuum also suggests a high peak accretion rate, although the range of viable values we find is wider, $10^{-8}~\mathrm{M_\odot~yr}^{-1} \lesssim \dot{M}_{\rm acc} \lesssim 10^{-7}~\mathrm{M_\odot~yr}^{-1}$. The reason for this is that the wind itself can also contribute significantly to the optical continuum flux, by reprocessing high-energy disk emission to longer wavelengths.

\citetalias{tam22v455andspec} analysed a set of spectra obtained during its 2007 outburst. They argued that the optical spectra of V455~And around the outburst maximum contained signatures of emission from a disc wind in the line profiles.
Figure \ref{fig:v455andspec} presents the low-resolution ($R\approx1400$) spectra observed at the Bisei Astronomical Observatory from \citetalias{tam22v455andspec}.
The spectrum around the outburst maximum (on BJD 2454350.04; top spectra in figure \ref{fig:v455andspec}) shows the strong (equivalent width (EW) of H$\alpha < -60$ \AA), broad-based (full width of zero intensity FWZI $\approx 5000$ km s$^{-1}$), but narrow-peaked (single-peaked in H$\alpha$) emission lines of Balmer and He~\textsc{ii}.
Considering the high-inclination nature of V455~And, these emission lines resemble those in SW Sex-type NLs.
The time-resolved spectra did not show any significant changes in radial velocities or eclipses over the orbital phases, suggesting that the emitting source cannot be related to the hotspot or the irradiated secondary star, but most likely originated in the disc wind.  
These spectral features are different from other DNe, especially from WZ Sge which is thought to share the high-inclination nature with V455~And.
WZ Sge at its outburst maximum showed the Balmer absorption lines and the double-peaked \heiifses~ with the peak separation of $\geq 1000$ km s$^{-1}$, implying the irradiated disc origin for the optical spectral lines \citep{bab02wzsgeletter, kuu02wzsge, nog04wzsgespec}.
Moreover, the trends of the spectral features in WZ Sge-type DNe, such as line EWs v.s. inclinations, support this idea (see \citealt{tam21seimeiCVspec} and references therein).

Spectra obtained later during the outburst plateau on BJD 2454356.08 and 2454363.03 (middle and bottom spectra in figure \ref{fig:v455andspec}) showed weaker and double-peaked Balmer and He~\textsc{ii} emission lines.
The typical peak separations of emission lines on BJD 2454356.08 and 2454363.03 were 600 and 900 km s$^{-1}$, respectively. 
At least the larger of these is compatible with the Keplerian velocity near the outer edge of the disc. 
Of course, this does not necessarily mean that these emission lines {\em are} formed in the disc.
Later, \citet{tov22v455andspec} confirmed these results on V455~And.

\section{Monte Carlo simulation of ionization and radiative transfer}
\label{sec:3}

In an attempt to model the spectrum of V455 And, we have used the Monte Carlo ionization and radiative transfer code \python
\footnote{Python is a collaborative open-source project available at \url{https://github.com/agnwinds/python}.}
~\citepalias{lon02python}. 
When presented with a kinematic prescription for a biconical wind arising from the surface of an accretion disc, this code can be used to calculate the ionization structure of the wind and then to simulate the output spectra for comparison to observations. 
Although \python~ was originally developed to simulate the UV spectra of high-state CVs (\citetalias{lon02python}; \citealt{noe10rwtriuxuma}), various improvements have made it possible to also model the optical spectra (\citetalias{mat15CVdiskwind}; \citealt{ini22j0714}). 
The code is intended to be fairly general, and \python\ has also been used for simulations of spectra of young-stellar objects \citep{sim05ysospec},  X-ray binaries \citep{kol23j1820wind}, AGN/quasars \citep{hig13balquasar, hig14agnwind, mat16quasarclumpy, mat17quasarspec, mat20agnwind}, and tidal disruption events \citep{par20tdewind, par22tdeopticalwind}.

\subsection{Calculation scheme}
\label{subsec:python}

\python~ consists of three main sections \citepalias{lon02python}.
The first section is for reading the user inputs and defining the wind geometry, kinematics, and simulation grids.
After this, the second section, the ionization calculation, estimates the wind ionization and temperature structure by flying the Monte Carlo energy packets (``$r$-packets'') through the defined wind geometry.
These ionization cycles are iterated until the wind structure gets thermally stabilized. 
In the third and final section, the spectrum calculation, the synthetic spectrum is obtained by determining the emissivity from each grid cell and correctly tracking/weighting its contribution to the observed flux for any given viewing angle. 
Our simulation setup has generally followed \citetalias{mat15CVdiskwind}, unless explicitly noted otherwise.

\subsection{Binary parameters of V455~And}
\label{subsec:binaryparam}

For the purposes of our calculations, we have used the binary parameters listed in Table \ref{tab:binaryparam}.
V455~And is a typical low-mass-ratio, short-orbital-period, and eclipsing WZ Sge-type DN (see Section \ref{sec:2} and \citetalias{ara05v455and}; \citealt{kat15wzsge}).
\citetalias{ara05v455and} arbitrarily applied the WD mass $M_{\rm WD}$ as 0.6 M$_\odot$, but no dynamical mass estimations have been performed so far on V455~And.
We fixed at $M_{\rm WD} =0.8$M$_\odot$ as \citet{tov22v455andspec}, which is the median value of the WD mass in ordinary CVs \citep{zor11SDSSCVWDmass, pal22WDinCVs}.
The WD radius $R_{\rm WD}$ is set as $7 \times 10^{8}$ cm, following the standard mass-radius relation in WDs \citep{fon01whitedwarf}.
\citetalias{ara05v455and}; \citet{szk13V455And} reported the WD temperature before and after the outburst at around 11000 K.
The lower limit of the WD temperature is given by this value, but as we model this system in outburst, the WD temperature should be hotter \citep{sio95CVWDheating, pir05wdcooling}.
Since the WD contributes very little to the overall emission from the system in the situations we are modelling, we fixed the WD temperature $T_{\rm WD}$ at 40000 K, also following \citetalias{lon02python, mat15CVdiskwind}.
The WD temperature of $\sim 40000$ K is observed in NLs \citep{tow09CVWDtemp} and expected in DN outbursts \citep{sio95CVWDheating}.
Based on the V455~And nature as a WZ Sge-type DN, the secondary mass $M_2$ and the mass ratio $q$ are set as 0.08M$_\odot$ and 0.1, respectively.
We assume the tidal truncation radius \citep{pac77ADmodel} to be the maximum disc radius $R_{\rm disc}$.
Applying the above WD mass, orbital period, and mass ratio we get $R_{\rm disc} = 2.2\times 10^{10}$ cm.
All the simulated spectra are calculated as the inclination $i=75^\circ$ \citepalias{ara05v455and} unless otherwise specified.

The other spectral synthesis calculations of high-state CVs using \python~ by \citetalias{lon02python, mat15CVdiskwind}; \citet{noe10rwtriuxuma, ini22j0714} assumed typical binary parameters for CVs above the so-called period gap (that is, P$_{\rm orb} \geq$ 3 hr and $q \geq$ 0.3; right column in table \ref{tab:binaryparam}).
We note that, although \citetalias{mat15CVdiskwind} originally adopted the inclination of RW Tri as $i=80^\circ$, we here present the results assuming $i=75^\circ$ for making a quantitative comparison to V455~And (see section \ref{subsec:nlmodels}).

\begin{table}
    \centering
    \caption{The adopted binary parameters for V455~And (this paper) and  RW Tri \citepalias{mat15CVdiskwind}.}
    \begin{tabular}{llcc} 
        \hline
             &  & V455~And & RW Tri \\
             &  & (this paper) & \citepalias{mat15CVdiskwind} \\
        \hline
            WD mass & $M_{\rm WD}$ [M$_\odot$]    & 0.8               & 0.8 \\
            WD radius & $R_{\rm WD}$ [cm]           & $7\times 10^{8}$  & $7\times 10^{8}$  \\
            WD temperature & $T_{\rm WD}$ [K]            & 40000             & 40000 \\
            Secondary mass & $M_{2}$      [M$_\odot$]    & 0.08 & 0.6                \\
            Mass ratio ($M_{2}$/$M_{\rm WD}$)& q          & 0.1                  & 0.75                \\
            Orbital period & $P_{\rm orb}$ [hr]          & 1.35     & 5.57              \\
            Inclination & $i$ [$^\circ$]           & 75         & 75              \\
            disc maximum radius & $R_{\rm disc}$ [cm]     & $2.2\times 10^{10}$    & $2.4\times 10^{10}$ \\
        \hline
    \end{tabular}
    \label{tab:binaryparam}
\end{table}

\subsection{Disc wind model}
\label{subsec:windmodel}

\begin{figure}
    \centering
    \includegraphics[width=0.7\linewidth]{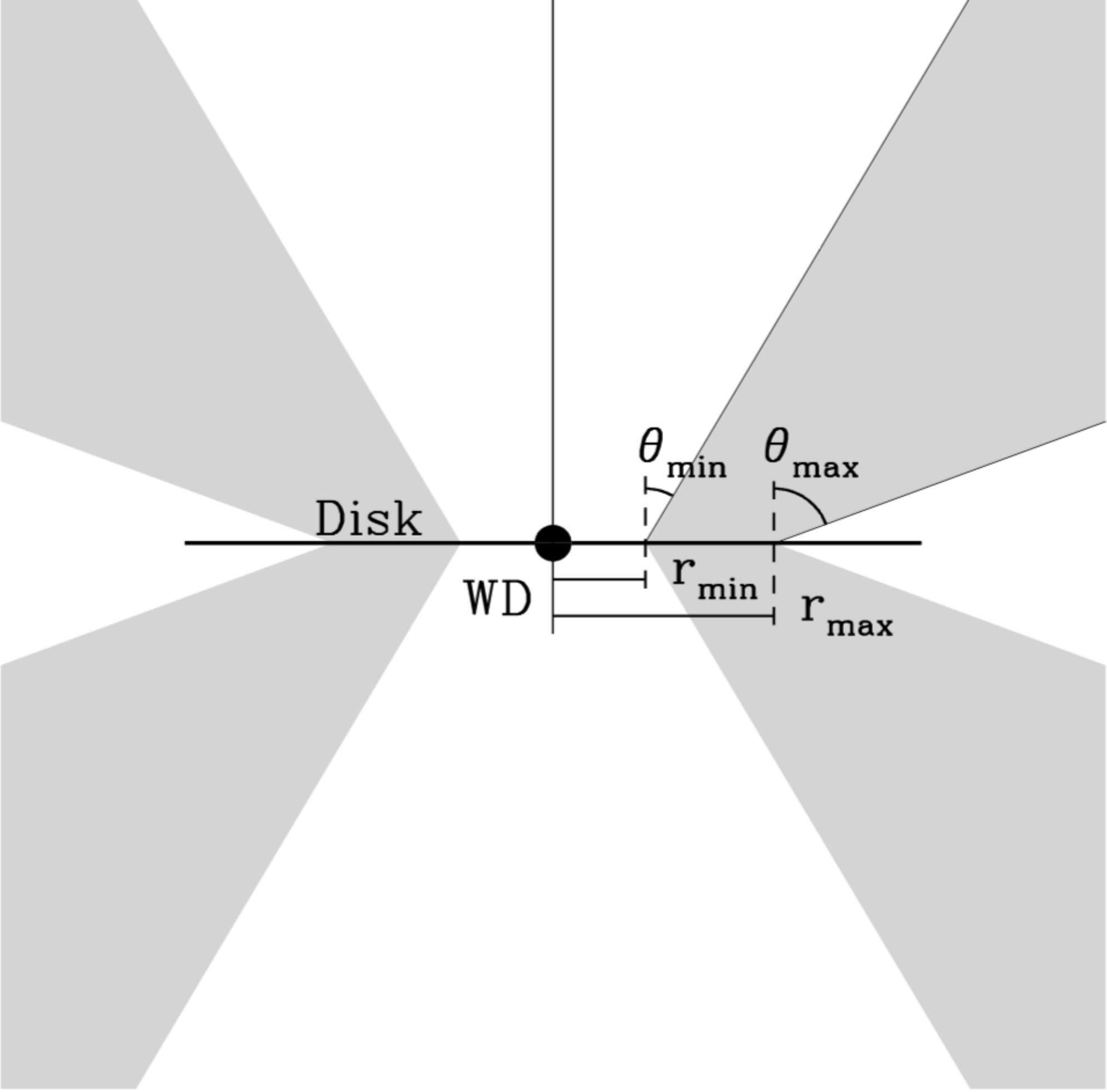}
    \caption{
    A cartoon showing the geometry and some key parameters of the biconical wind model by \citet{shl93CVwind}.
    }
    \label{fig:geometry}
\end{figure}

\subsubsection{Geometry and kinematics}

For this study, we have adopted the prescription for a smooth biconical outflow first proposed by \citet{shl93CVwind}. 
This prescription has been used in several previous studies as it is physically plausible, relatively flexible, and has a reasonably small number of parameters. 
The basic wind geometry is shown in Figure \ref{fig:geometry}.
The winds emanate from a disc between the radii $r_{\rm min}$ and $r_{\rm max}$, and its inner and outer opening angles are $\theta_{\rm min}$ and $\theta_{\rm max}$, respectively.
The launch angle of a wind streamline from a launch radius $r_0$ (where r$_{\rm min} \leqq r_0 \leqq r_{\rm max}$) is given by 

\begin{equation}
    \theta\left({r_0}\right) = \theta_{\rm min} + \left( \theta_{\rm max} - \theta_{\rm min} \right)
    \left( \frac{r_{0} - r_{\rm min}}{r_{\rm max} - r_{\rm min}} \right)^\gamma ,
	\label{eq:thetalaw}
\end{equation}
where the exponent $\gamma$ controls the concentration of the streamlines; in this work we set $\gamma=1$, corresponding to uniform angular spacing. 
The poloidal velocity field $v_l$ along the wind streamline from $r_0$ is given by
\begin{equation}
    v_l = v_0 + \left(v_\infty(r_0) - v_0 \right) \frac{\left(l/R_v\right)^\alpha}{\left(l/R_v\right)^\alpha + 1},
	\label{eq:vellaw}
\end{equation}
where $v_0$ is the launch velocity and $v_\infty(r_0)$ is the terminal velocity.
$R_v$ and $\alpha$ are the acceleration length and the acceleration exponent, respectively, which control the wind acceleration along the streamline. 
The rotational velocity $v_\phi$ is determined to conserve the specific angular momentum of the initial Keplerian rotation at $r_0$ around the primary WD, such that
\begin{equation}
    v_\phi = \left(\frac{G M_{\rm WD}}{r_0}\right)^{1/2} \frac{r_0}{r}.
	\label{eq:vphi}
\end{equation}
We fixed $r_{\rm min}$ = 4 $R_{\rm WD}$,  $r_{\rm max}$ = 12 $R_{\rm WD}$, $\theta_{\rm min}$ = 20$^\circ$,  $\theta_{\rm max}$ = 65$^\circ$, $v_0$ = 6 km s$^{-1}$ and $v_\infty(r_0) = 3 v_{\rm esc}(r_0)$, where $v_{\rm esc}(r_0)$ is the escape velocity at $r_0$, for all the calculations presented in this article, which are the same values adopted by \citet[][also see \citetalias{lon02python, mat15CVdiskwind}]{shl93CVwind}.
These parameters have relatively minor impacts on the resulting spectra, compared to the parameters we allow to vary: accretion rate $\dot{M}_{\rm acc}$, wind mass-loss rate $\dot{M}_{\rm wind}$, volume filling factor $f_V$, acceleration length $R_V$ and exponent $\alpha$.

\begin{table*}
    \centering
    \caption{\python~ parameters describing the geometry and kinematics of the wind.
    Models A, B X are taken from \citetalias{lon02python,mat15CVdiskwind}.
    The other two models, Models I and II, are optimized to explain the V455~And spectrum (see section \ref{sec:4}).}
    \begin{tabular}{rc|ccccccc} 
        \hline
             &        &  \multicolumn{5}{c}{Model} \\
             &        &  A &  B &  X &  I &  II \\
             \multicolumn{2}{c}{Parameters} & \citetalias{lon02python} & \citetalias{mat15CVdiskwind} & \citetalias{mat15CVdiskwind} & this paper & this paper\\
            \hline
            disc accretion rate & $\dot{M}_{\rm acc}$ [M$_\odot$ yr$^{-1}$] & $1\times 10^{-8}$ & $1\times 10^{-8}$ & $1\times 10^{-8}$ & $5\times 10^{-8}$ &  $1\times 10^{-7}$\\
            wind mass-loss rate & $\dot{M}_{\rm wind}$ [M$_\odot$ yr$^{-1}$] & $1\times 10^{-9}$ & $1\times 10^{-9}$ & $1\times 10^{-9}$ & $2\times 10^{-8}$ & $1\times 10^{-8}$ \\
            acceleration length & $R_v$ [cm] & 7 $\times$ 10$^{10}$ & 1 $\times$ 10$^{11}$ & 2 $\times$ 10$^{11}$ & 7 $\times$ 10$^{10}$ & 7 $\times$ 10$^{10}$ \\
            acceleration exponent & $\alpha$                      & 1.5 & 4 & 4 & 1.5 & 1.5 \\
            Volume filling factor  & $f_V$        & 1 & 1 & 1 & 1 & 0.1 \\
            BL luminosity  & $L_{\rm BL}$ [erg s$^{-1}$]     & -- & -- & -- & $1.1\times10^{35}$ & $2.2\times10^{35}$ \\
            BL temperature & $T_{\rm BL}$ [K] & -- & -- & -- & 10$^5$ & 10$^5$ \\
        \hline
	\end{tabular}
    \label{tab:windparam}
\end{table*}

Here, we present the results of five wind models.
Table \ref{tab:windparam} presents the summary of the wind parameters.
The first three wind models (the reference wind models) are the ones from \citetalias{lon02python, mat15CVdiskwind}, while the binary parameters are updated for the ones of V455~And in this paper (see table \ref{tab:binaryparam}).
Model A was devoted to reproducing the UV spectrum of Z Cam in \citetalias{lon02python}, while Model B and X are more focused on reproducing the optical spectra of RW Tri \citepalias{mat15CVdiskwind}.
The other two models, Models I and II, are introduced in this paper to explain the observed spectrum of V455~And around its outburst maximum.
As can be seen in table \ref{tab:windparam}, we applied higher disc accretion rates in both Models I and II than those of the reference models, following the expected accretion rate using the binary parameter of V455~And (see section \ref{sec:2}).

We note that we have calculated an extensive grid of models exploring different wind acceleration parameter sets and ratios of mass-loss to accretion rates assuming the wind mass-loss rate $\dot{M}_{\rm wind} = 10^{-9}$ M$_\odot$ yr$^{-1}$; none of these reproduced the observations. 
In all of our tests, the only part of parameter space capable of reconciling the simulations with the observations corresponds to models with wind mass-loss rates $\dot{M}_{\rm wind} \approx 10^{-8}$ M$_\odot$ yr$^{-1}$.
The summary of this parameter survey will be published as a subsequent paper (Tampo et. al., in prep.).

\subsubsection{Atomic data}

Following \citet{sim05ysospec} and \citetalias{mat15CVdiskwind}, we applied the hybrid approach to calculate the ionization and excitation state of the atoms; the macro-atom model for H and He  \citep{luc02macroatomI, luc03macroatomII}, but the simple (two-level) atom model for other metals.
The atomic data are the same as the calculations in \citetalias{mat15CVdiskwind}, 
which has been adopted from \citet{hig13balquasar} for the metals, from \citet{sim05ysospec} for H, and from \citet{bad05opacitymodel} for He.

\subsubsection{Clumping of winds}

We also allow the wind to have a clumpy structure.
In terms of observations, variable absorption components of P-Cygni profiles in UV (e.g., \citealt{woo90suumarxandbzcamUV, pri00bzcamwindHST}) and in optical \citep{kaf04windfromCV, cun23nloutflow}, as well as time-variable absorbers detected in X-rays \citep{sai12zcam, dut23sscygxray},  suggest the presence of clumpy structures in CV disc winds.

\python\ contains a simple version of clumping, called 'microclumping' \citep{mat16quasarclumpy}.
This microclumping technique was originally developed for the modelling of stellar winds to reproduce strong emission lines while keeping a low mass-loss rate \citep{ham95wrstarclump, ham98stellarwind, hil99hd165763}.
In this scheme, the density is enhanced by the factor $D$, while the volume filling factor is reduced by $f_V = D^{-1}$.
The clump sizes are assumed to be much smaller than the typical photon mean free path and thus the inter-clump medium does not affect the output spectrum. 
The result of this approach is that any physical process linear in density is unaffected, while any processes scaling the density square (i.e., collisional excitation, recombination) are enhanced by a factor of $D$.

\subsection{Sources and sinks of radiation}
\label{subsec:sinks}

As sources of radiation, we consider the accretion disc, primary WD, and BL.  In the models calculated here,  we treat the disc and WD as purely reflecting.
Our model does not include emissions from either the hotspot or the secondary star.
This is justified by the low mass-transfer-rate and low-mass-ratio nature of V455~And.
However, \python~ can calculate an eclipsed spectrum by the Roche-lobe-filling secondary star, 
which requires one to specify the orbital period, $P_{\rm orb}$, secondary mass, $M_2$, and the orbital phase $\phi_{\rm orb}$.

\subsubsection{Accretion disc}
\label{sec:341}

\python~ is capable of calculating the emission from an accretion disc as either a multi-temperature blackbody standard disc model \citep{sha73alphadisk} or a stellar atmosphere model with the appropriate surface gravity and temperature. The disc follows the Keplerian rotation around the primary WD.
When spectra from the stellar atmospheres are used, the spectra have been calculated using {\textsc{synspec}\footnote{\url{http://nova.astro.umd.edu/Synspec43/synspec.html}} based on stellar atmospheres obtained 
from either \citet{kur91stellaratmospheres} ($T_{\rm eff} \leq$ 50000 K) or calculated with \textsc{tlusty} ($T_{\rm eff} >$ 50000 K; \citealt{hub95nonlte}).
We break down the disc into multiple annuli such that each annulus contributes an equal amount to the bolometric luminosity. 
Following \citetalias{mat15CVdiskwind}, we use a standard geometrically-thin disc with a \cite{sha73alphadisk} temperature profile, with each annulus treated as a blackbody for the ionization cycles, but with stellar atmosphere models for the spectral cycles to obtain more realistic spectra. 
}

\subsubsection{White dwarf}
\label{sec:342}

The WD is set to be located in the centre of the disc.
We include the WD as a blackbody radiation source with temperature $T_{\rm WD}$ and radius $R_{\rm WD}$.
In our models, we fixed $T_{\rm WD} = 40000$ K and  $R_{\rm WD} = 7\times10^8$ cm (table \ref{tab:binaryparam}).

\subsubsection{Boundary layer}
\label{sec:343}

We have included a BL in Models I and II, mainly to account for the optical He~\textsc{ii} lines in the spectra during the outburst plateau phase (on BJD 2454356.08 and 2454363.03; see section \ref{subsec:timeevol}). 
The formation of these lines typically requires an EUV-bright SED component, but at these later times, $\dot{M}_{\rm acc} \lesssim 5 \times 10^{-9}~\mathrm{M_\odot~yr}^{-1}$, and the maximum disk temperature is $\leq 60000$ K.

The BL in high-state CVs is generally thought to be optically thick (e.g., \citealt{pri77softXinDN, mau02CVinEUVE}).
Therefore, we treat the BL as a point source of blackbody radiation located at the centre of the disc with temperature $T_{\rm BL} = 10^5$~K and luminosity $L_{\rm BL}$ equal to the bolometric luminosity of the accretion disc (see Table \ref{tab:binaryparam}). 

There is only report of X-ray observations obtained during the 2007 outburst (\citealt{sen08v455andhardX} = ATel \#1372). The BAT instrument on {\em Swift} detected the system in hard X-rays (14  -- 25 keV), with a peak flux corresponding to $L_{14-25~{\rm keV}} \simeq 7 \times 10^{31}~{\rm erg}~{\rm s}^{-1}$ at the {\em Gaia} parallax distance. This peak was seen on BJD 2454358, roughly one week after our first observation, which is the main focus of our paper. Even this peak hard X-ray luminosity corresponds to only $\simeq$ 0.3\% of the disk luminosity at the time of our first observation. We therefore neglect this component in our modelling.

V455~And was also detected as a bright (0.77 count s$^{-1}$) {\em soft} X-ray source by the XRT instrument on {\em Swift} during the outburst, around on BJD 2454361, roughly 11 d after our first observation \citep{sen08v455andhardX}. It is therefore interesting to ask if this is compatible with the boundary layer in our models. Unfortunately, no direct estimate of the neutral hydrogen column density along the line of sight to the system, $N_H$, is available. However, the nearby distance and the lack of any obvious interstellar Lyman $\alpha$ absorption in the far-ultraviolet spectra \citep{ara05v455and, szk13V455And} suggest that $N_H$ is low. 

In order to arrive at a quantitative estimate, we note that the distance of V455 And is almost identical to that of VW Hyi, and the total Galactic extinction along both lines of sight is also virtually identical (corresponding to $E_{B-V} \simeq 0.095$). If we assume that $N_H$ along both lines of sight is comparable as well, we can adopt the well-determined column density for VW~Hyi ($N_H \simeq 6 \times 10^{17}$~cm$^{-2}$; \citealt{pol90cveuv}) as an estimate for V455~And also. Using WebPIMMs\footnote{\url{https://heasarc.gsfc.nasa.gov/cgi-bin/Tools/w3pimms/w3pimms.pl}}, we then find that the observed count rate corresponds to $L_{\rm BL} \simeq 2 \times 10^{35}$~erg~s$^{-1}$ for our adopted $T_{\rm BL} \simeq 10^5$~K. This is in good agreement with our model for the first observation (BJD 2454350), though significantly brighter than we assume in Section~\ref{subsec:timeevol} for BJD 2454363 near the end of the outburst. However, this inference is subject to significant systematic uncertainties. For example, the BL luminosity we assume for BJD 2454363 ($L_{\rm BL} \simeq 10^{33}$~erg~s$^{-1}$) becomes consistent with the {\em Swift}/XRT count rate observed on BJD 2454361 if the BL temperature is changed to a still reasonable value of $T_{\rm BL} \simeq 2 \times 10^5$~K. 
Unfortunately, there is no soft X-ray or UV information in this ATel during the earlier part of the outburst to constrain the evolution of the inner disk and BL luminosity. 
Although the powerlaw component detected in hard X-rays can contribute to a part of the soft X-ray flux, as long as the total EUV/soft X-ray flux from multiple radiation sources is at the same (or lower) level as the optically-thick BL model, simulated optical spectra would be similar.

It should finally be acknowledged that, although there are some unambiguous observational examples of optically-thick BL (e.g., \citealt{lon96ueveugem, kim23j0302}), there is still significant debate regarding the existence/properties of BLs in high-state CVs \cite[e.g][]{muk17xrayawd}. We therefore note that including the BL does not make any significant difference for the overall output spectra for Models I and II near outburst maximum. In our models for this epoch, the disk accretes rapidly enough and is hot enough to produce sufficient (E)UV radiation even without a BL contribution. We note in passing that adding such a component may also provide a rough way to allow for the possibility that the WD in V455~And undergoes steady nuclear burning near outburst maximum, given the very high accretion rate implied by the optical brightness of the source (c.f. Section \ref{sec:2}). Of course, if the system really becomes a steady-burning supersoft X-ray source near outburst maximum, the soft X-ray component could be significantly stronger than in our models.

\section{Results}
\label{sec:4}

In this section, we first compare the observed and synthesized spectra from the reference wind models (Models A, B, and X) to check what respects these models can and cannot explain the observations of V455~And.
Then we present the models (Models I and II) tuned to match the observations.

\subsection{Reference wind models}
\label{subsec:nlmodels}

\begin{table*}
    \centering
    \caption{The observed and simulated line flux and EW of H$\alpha$, H$\beta$ and \heiifses.
    The orbital parameters of the simulated spectra are set at $D=$75.6 pc, $i=75^\circ$, and $\phi_{\rm orb}=0.5$.
    Since the EW is defined for absorption lines, emission lines give negative EWs.}
    \begin{tabular}{l|cc|cc|cc} 
        \hline
            & \multicolumn{2}{c}{H$\alpha$} & \multicolumn{2}{c}{H$\beta$} & \multicolumn{2}{c}{\heiifses} \\
            & EW [\AA] & Flux [erg s$^{-1}$ cm$^{-2}$] & EW [\AA] & Flux [erg s$^{-1}$ cm$^{-2}$] & EW [\AA] & Flux [erg s$^{-1}$ cm$^{-2}$] \\
        \hline
        V455~And &  $-$68.1(1.4) & 3.85(3) $\times 10^{-11}$ & 
                    $-$25.2(6) & 2.68(2) $\times 10^{-11}$ & 
                    $-$27.0(3)$^*$ & 5.39(3) $\times 10^{-11}$$^*$ \\
        \hline
        Model A &   $-$15.4(1) & 2.36(1) $\times 10^{-12}$ & 
                    $-$2.6(1) & 1.28(3) $\times 10^{-12}$ & 
                    $-$8.1(1)  & 3.82(3) $\times 10^{-12}$ \\
        Model B &   $-$19.4(2) & 8.55(3) $\times 10^{-12}$ & 
                    $-$9.5(1) & 1.05(1) $\times 10^{-11}$ & 
                    $-$10.3(1) & 1.34(1) $\times 10^{-11}$\\
        Model X &   $-$20.6(2) & 1.31(1) $\times 10^{-11}$ & 
                    $-$10.4(1) & 1.49(2) $\times 10^{-11}$ & 
                    $-$8.0(1) & 1.45(2) $\times 10^{-11}$\\
        Model I &   $-$67.8(2) & 3.18(1) $\times 10^{-11}$ & 
                    $-$17.0(1) & 2.52(1) $\times 10^{-11}$ & 
                    $-$25.2(2) & 4.04(2) $\times 10^{-11}$\\
        Model II &  $-$80.0(3) & 4.83(2) $\times 10^{-11}$& 
                    $-$21.8(2) & 4.11(2) $\times 10^{-11}$ & 
                    $-$31.5(2) & 6.47(3) $\times 10^{-11}$\\
        \hline
        \multicolumn{7}{l}{$^*$this value can be contaminated with the Bowen blend and He~\textsc{i} $\lambda$4713.}\\
	\end{tabular}
    \label{tab:specflux}
\end{table*}

\begin{figure}
    \centering
    \includegraphics[width=0.98\linewidth]{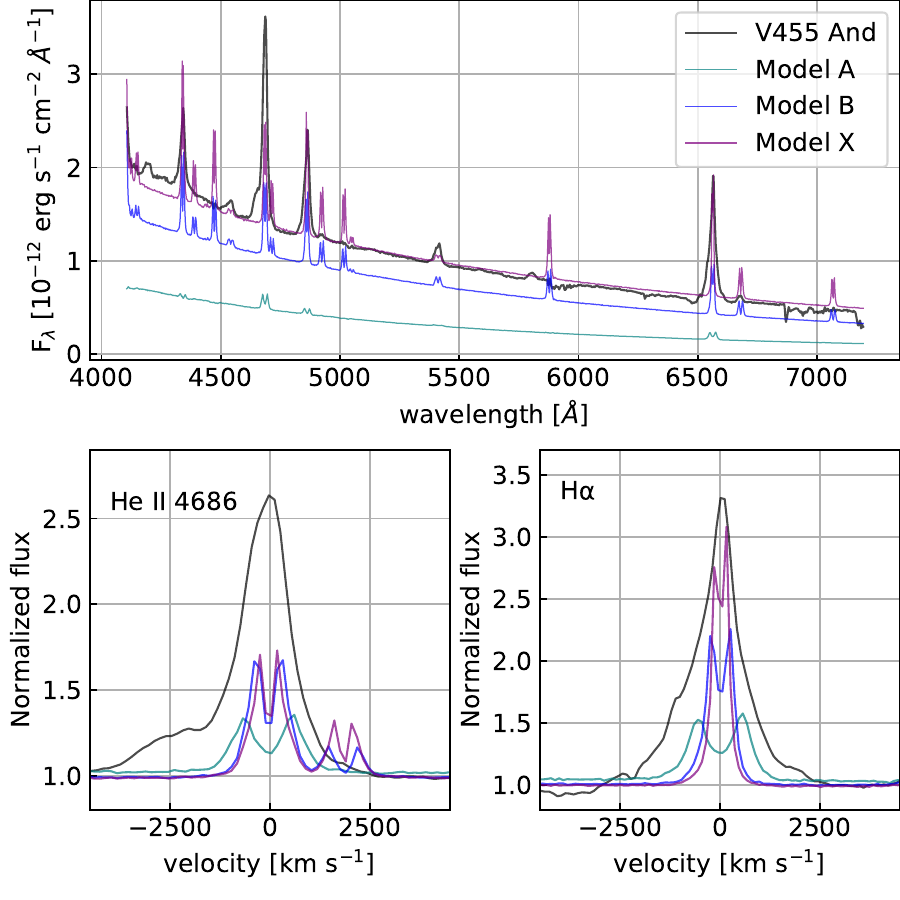}
    \caption{
    The flux-calibrated spectrum (top panel) and normalized line profiles of H$\alpha$ (right-bottom) and \heiifses~ (left-bottom).
    The black lines represent the V455~And spectrum around the outburst maximum (BJD 2454350.04), while the cyan, blue, and purple lines show the simulated results of Models A, B, and X, respectively.
    }
    \label{fig:jm15models}
\end{figure}

In figure \ref{fig:jm15models},  we show the simulated spectra from models that use the same wind parameters as \citetalias{lon02python, mat15CVdiskwind} (Models A, B, and X).
The black lines represent the optical spectrum of V455~And observed on BJD 2454350.04 (i.e., around the outburst maximum).
The spectral resolution of these observations is $R \sim 1400$ and the spectrum presented in this figure is outside of the eclipse ($\phi_{\rm orb}\approx0.7$).
The top panel shows the overall flux-calibrated spectrum and the bottom panels show the line profile of H$\alpha$ (right) and \heiifses~ (right) normalized by the continuum. 
The coloured lines present the simulated spectra smoothed to the spectral resolution of the observations (cyan; Model A, blue; Model B, purple: Model X from \citetalias{mat15CVdiskwind}) at $D=$75.6 pc, $i=75^\circ$, and $\phi_{\rm orb}=0.5$.
We note that a bump seen around 4650\AA~ ($-2500$ km s$^{-1}$ in the bottom left panel) is the C~\textsc{iii} /  N~\textsc{iii}  Bowen blend, which is not included in our atomic model.
Table \ref{tab:specflux} summarizes the observed and the simulated line fluxes and EWs of H$\alpha$, H$\beta$, and \heiifses.

Although the simulated spectra show the strong emission lines of Balmer, He~\textsc{i}, and He~\textsc{ii}, 
their profiles are not strong enough to explain the observations.
Moreover, only Model X reproduces the reasonable fit to the continuum level.
We here summarise the four points that these reference models do not match with the observations; 
the models show 
(\textsc{1}) the weaker emission lines than the observations both in terms of line fluxes and EWs (the simulated H$\alpha$ EWs are $>-25$ \AA~ while that in the observation is $<-60$ \AA), 
(\textsc{2}) the narrow line profiles at the base with the FWZI of $\leq$ 3000 km s$^{-1}$ compared to the observations with the FWZI of $\geq$ 4000 km s$^{-1}$,  
(\textsc{3}) the smaller line flux and EW ratios of \heiifses~ over H$\beta$,
and
(\textsc{4}) the double-peaked emission profiles with a clear central dip.

\begin{figure}
    \centering
    \includegraphics[width=0.8\linewidth]{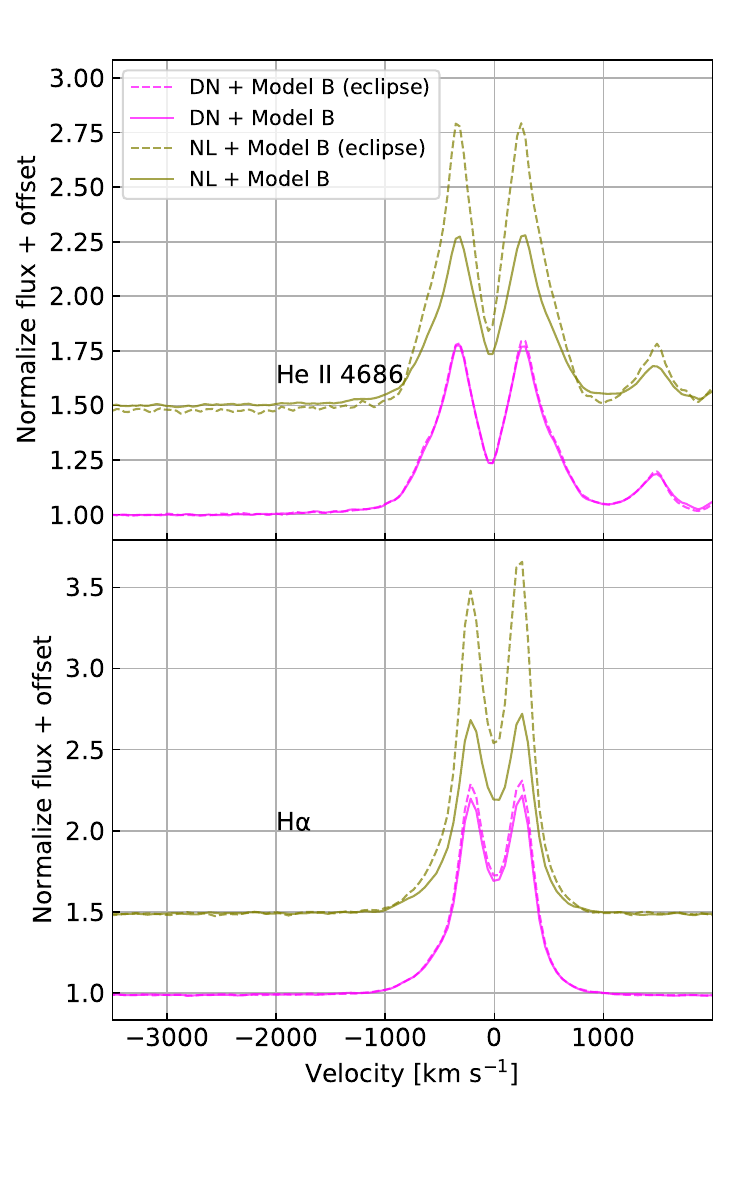}
    \caption{The line profiles of \heiifses~ (top) and H$\alpha$ (bottom).
    The lines show the model using the same Model B wind parameters at the same inclination of 75$^\circ$ but using the different binary mass ratio of V455~And ($q=0.1$; {magenta}) and RW Tri ($q=0.75$; {green}), respectively.
    The solid and dashed lines represent the line profile at orbital phase $\phi_{\rm orb}=0.5$ (i.e., out of eclipse) and orbital phase $\phi_{\rm orb}=0.0$ (i.e., primary eclipse).
    }
    \label{fig:eclipse}
\end{figure}

In figure \ref{fig:eclipse}, we compare the effects of the primary eclipse on the spectra for different mass ratios. 
This is effectively a comparison between the spectra expected for CVs with short ($\leq$ 2 hr) and long ($\geq$ 3 hr)  orbital periods. 
We focus on Model B here, comparing results for the binary parameters of V455~And (DN: $q=0.1$ and $i=75^\circ$; magenta) and RW Tri (NL: $q=0.75$ and $i=75^\circ$; green; \citetalias{mat15CVdiskwind}; see right column in table \ref{tab:binaryparam}).
The solid lines show the line profile at phase $\phi_{\rm orb}=0.5$ (i.e. outside of the eclipse), while the dashed lines show the line profile at phase $\phi_{\rm orb}=0.0$ (i.e. in primary eclipse).
By comparing the solid lines, which stand for the out-of-eclipse epochs, it is clear that the change of the binary parameter does not have any major effect on the line profiles.
However, the eclipse has a greater impact on the line profile in the $q=0.75$ case due to the larger size of the secondary star.
In the $q=0.1$ case, the H$\alpha$ EW increases less than 10$\%$ between inside and outside of the eclipse, while this increase is $\sim 70\%$ for the $q=0.75$ case.
In observations, V455~And did not show any strong evidence of an eclipse on its line profiles (\citetalias{tam22v455andspec}; \citealt{tov22v455andspec}). 
On the other hand, the increase of EWs around the primary eclipses is commonly observed in the time-resolved spectra of SW Sex-type NLs (e.g., \citealt{rod01v348pup}).
Therefore, in the disc wind model, the different behaviour of the time-resolved spectra around the primary eclipses are simply understood as the different sizes of the secondary star between WZ Sge-type DNe and SW Sex-type NLs.
Future observations of eclipsing systems with various mass ratios may benefit understanding the wind structure.

\subsection{Custom disc wind models: matching V455~And with dense and/or clumpy outflows}
\label{subsec:bestmodels}

Here, we present two models (Models I and II; table \ref{tab:windparam}) that successfully reproduced the spectral features of V455~And around the outburst maximum.
Model I is characterized by its high wind mass-loss rate, reaching 40\% of the underlying disc accretion rate.
This mass-loss to accretion rate ratio is remarkably higher than the disc wind models applied for UV spectra in other high-state CVs (1 -- 10\%; e.g., \citealt{vit93cvwind}; \citetalias{lon02python}).
Model II has a relatively normal mass-loss rate to accretion rate ratio of 10\%, but is characterized by the clumpy wind structure with the volume filling factor $f_V =$ 0.1.
\citet{ham98stellarwind} noted that the microclumping scheme tends to reduce the mass-loss rate required to wind-formed lines by a factor of $\sqrt{D}$, which is comparable to the difference in the mass-loss to accretion rates ratios between Models I and II. Indeed, $f_V \simeq 0.1$ is often adopted in stellar winds (e.g., \citealt{hil99hd165763}).
Meanwhile, both in Models I and II, we applied the wind acceleration law the same as in Model I.

\begin{figure*}
    \centering
    \includegraphics[width=0.49\linewidth]{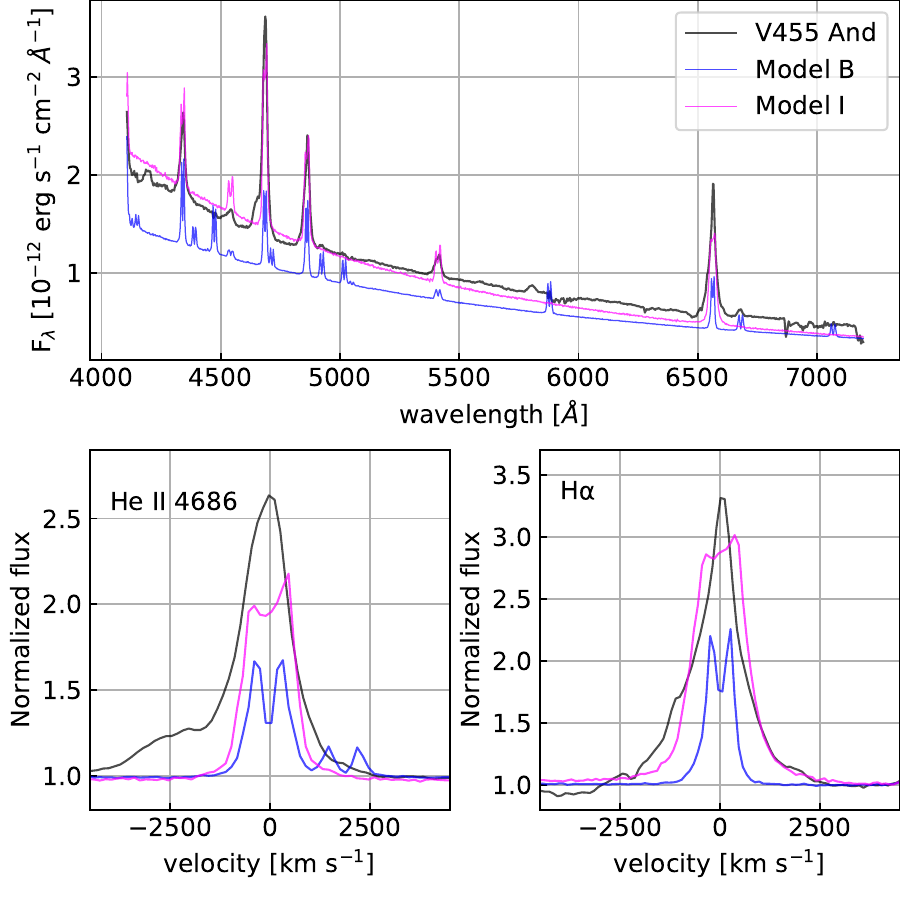}
    \includegraphics[width=0.49\linewidth]{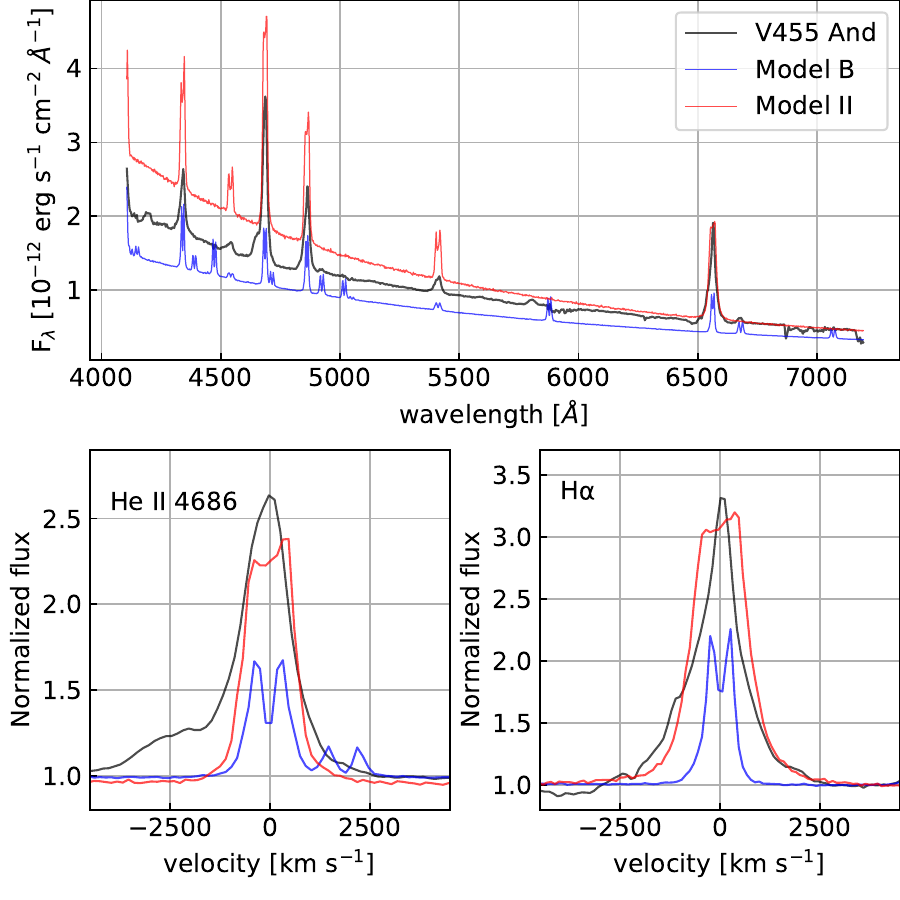}
    \caption{Same figures as figure \ref{fig:jm15models}, but for Model I (left panels; magenta) and Model II (right panels; red).}
    \label{fig:modelspec}
\end{figure*}

Figure \ref{fig:modelspec} presents the simulated spectra of Model I (magenta on the left panel) and Model II (red on the right panel), along with the observations (black) and Model B (blue) for comparison.
The spectra of Models I and II show a series of strong emission lines associated with H~\textsc{i} and He~\textsc{ii}, while the He~\textsc{i} lines are relatively weak, as observed.
The line fluxes and EWs of the Balmer and \heiifses\ lines in both Models I and II are within a factor of 1.6 from the observed values (table \ref{tab:specflux}).
The Balmer FWZIs reach $\sim 5000$ km s$^{-1}$, as observed, while those of \heiifses~ are $\sim 4000$ km s$^{-1}$, slightly narrower than the observed values.
We note that the observed FWZI of \heiifses~ can be affected by the Bowen blend (around 4650\AA) and He~\textsc{i} $\lambda$4713 though.
On the other hand, both Models I and II show flat-top emission line profiles rather than the single-peaked profiles in the observations.
The dips at the line centres of Models I and II are much shallower than those in Models A, B, and X.

\begin{figure*}
    \centering
    \includegraphics[width=0.49\linewidth]{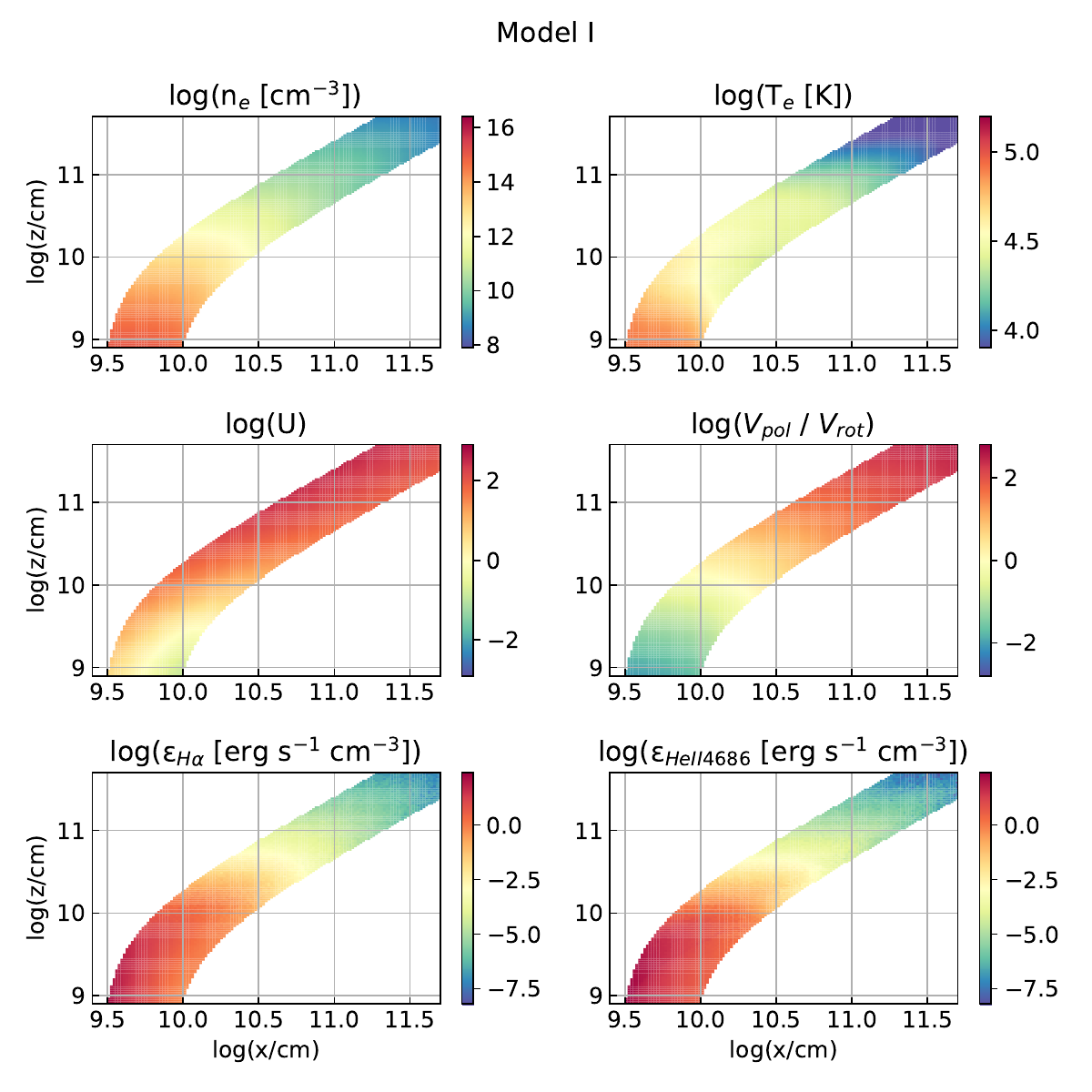}
    \includegraphics[width=0.49\linewidth]{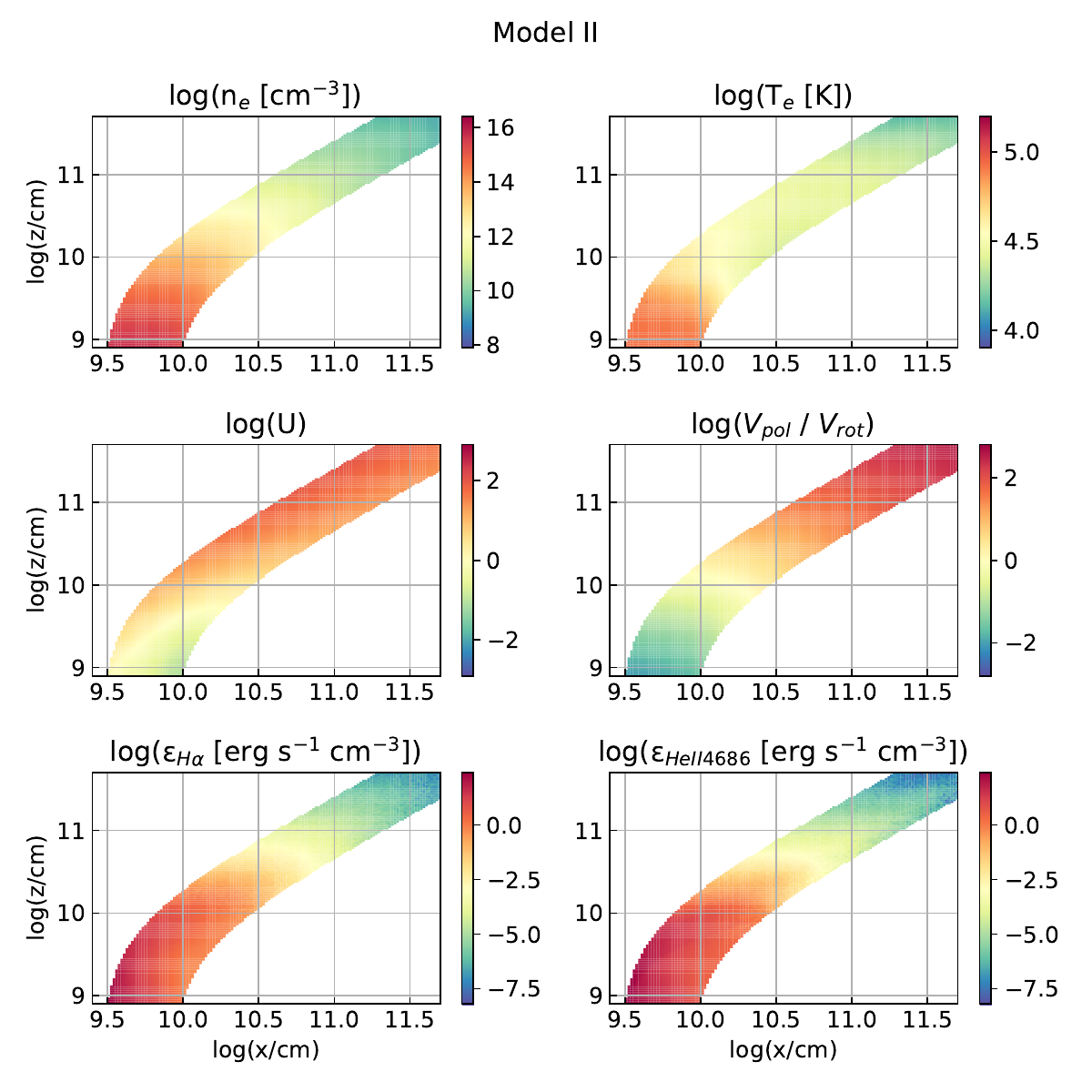}
    \caption{
    Wind structure of Model I (left) and Model II (right).
    The upper-left, upper-right, middle-left, middle-right, lower-left, and lower-right panels show the structure of electron number density $n_e$, electron temperature $T_e$, ionization parameter $U$, poloidal velocity $V_{\rm pol}$ over rotational velocity $V_{\rm rot}$, H$\alpha$ emissivity $\epsilon_{H\alpha}$ and \heiifses~ emissivity $\epsilon_{\rm He~\textsc{ii} \lambda4686}$, respectively.
    }
    \label{fig:modelmap}
\end{figure*}

Figure \ref{fig:modelmap} shows the kinematic and ionization structure of the wind of Model I (left) and Model II (right).
The ionization parameter $U$ represents the ratio of the number density of ionizing photons (that is, photons with $h \nu \geq 13.6~{\rm eV}$) to the local hydrogen density (equation 13 of \citetalias{mat15CVdiskwind}). It is calculated based on {\em all} of the ionizing photons passing through a given cell, although in practice the disc component dominates the luminosity near the H-ionizing threshold.

Because of the intrinsically increased mass-loss rates, the electron densities of the disc winds in Models I and II are indeed an order of magnitude higher than those in Models A, B, and X (also see figure 5 in \citetalias{mat15CVdiskwind}).
Moreover, the gentler acceleration (smaller acceleration exponent) in Models I and II than in Model B governs the vertically-thicker dense region below $R_V$, which results in the broader FWZI.  
The overall result is a larger line-forming region with low ionization parameters, higher line emissivity, and stronger emission lines in terms of both line fluxes and EWs in Models I and II.

We note in passing that the \ha:\hb\  line flux ratio observed in V455 And of 1.44(2) is considerably less than the value of 2.86 for recombination in the low-density limit \citep{ost06book}  (see also the right-handed panel of figure \ref{fig:spec-evol}).  Both the reference models and the custom models presented here show also show lower ratios, 1.26(1) and 1.18(1) for Models I and II, respectively.  These low ratios or ``flat Balmer decrements'' are expected, if as is the case, most of the line flux arises from regions of high density ($\gtrsim 10^{12}~{\rm cm}^{-3}$), as is discussed for the case of the X-ray binary MAXI J1820$+$070 by \citet[][see their Figure 9]{kol23j1820wind}.

\begin{figure}
    \centering
    \includegraphics[width=\linewidth]{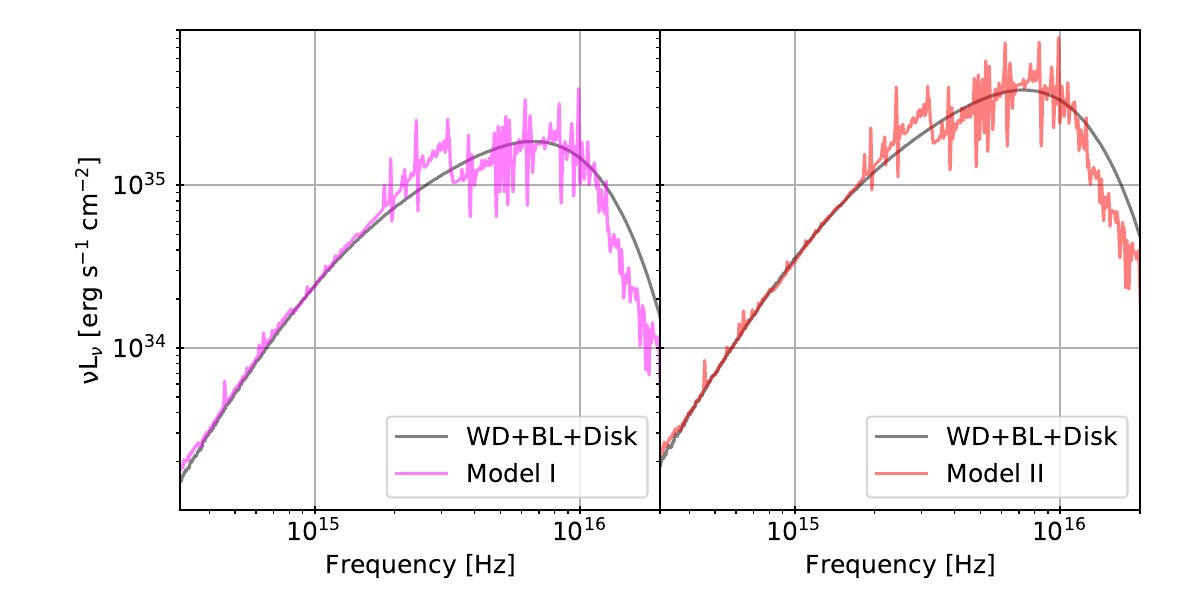}
    \caption{The packet-integrated spectra for all viewing angles of Model I (left) and II (right).
    The black lines represent the total input spectrum (the sum of the WD, BL, and disc), while the coloured lines show the output spectra.}
    \label{fig:modelsed}
\end{figure}

The broad-band angle-averaged spectral energy distributions are presented in figure \ref{fig:modelsed} for Model I (left) and Model II (right).
The black lines represent the input spectra composed of the primary WD, BL, and accretion disc.
These input emissions are absorbed and re-emitted through the disc wind and then integrated as the output spectra (Model I; magenta, Model II; red).
The wind absorbs photons efficiently at the Lyman and He~\textsc{ii} edges (912 and 228\AA, respectively), 
and then strongly reprocesses and re-emits in the UV--optical wavelengths, especially in recombination lines.

\section{Discussion}
\label{sec:5}

\subsection{The need for a high mass-loss rate and/or a clumpy disc wind in V455~And}
\label{subsec:windcond}

Our results show that to explain the observed spectral features of V455~And, achieving a high electron density via either a high wind mass-loss rate or a clumpy wind structure is crucial.
The required wind mass-loss rate in the V455~And case ($\dot{M}_{\rm wind} \simeq 10^{-8}$ M$_{\odot}$ yr $^{-1}$; Models I and II) is intrinsically an order or more of magnitude higher than in the 
reference models and in other literature ($\dot{M}_{\rm wind} \lesssim 10^{-9}$ M$_{\odot}$ yr $^{-1}$; Models A, B, and X). 
This comparatively high $\dot{M}_{\rm wind}$ may be reasonable, at least in a relative sense: as discussed in Sections~\ref{sec:2} and ~\ref{sec:3}, the accretion rate at outburst maximum in V455 And may be close to $10^{-7}$ M$_{\odot}$ yr $^{-1}$ (compared to $\sim 10^{-8}$ M$_{\odot}$ yr $^{-1}$ in typical high-state CVs). 
Line-driven wind models predict that the wind mass-loss rate is roughly proportional to the disc accretion rate (\citealt{pro99windcomparison} and references therein), and in a magneto-centrifugal wind, for fixed properties of the magnetic lever arm, a linear $\dot{M}_{\rm wind}\propto \dot{M}_{\rm acc}$ relationship is also expected \citep{pud07outflows}. 
The intrinsically high accretion rate in V455~And could be due to the low mass-transfer rate nature of DN outbursts (i.e., Z Cam-type DNe show brighter outburst maxima than standstills; \citealt{ste01zcamstandstill}), the additional angular momentum loss through tidal instability  \citep{lin79lowqdisk, osa02wzsgehump},  and/or a very low disc viscosity in quiescence in WZ Sge-type DNe, which naturally leads to higher accretion rates at outburst maximum \citep{min85DNDI, osa95wzsge}.

It therefore seems reasonable that V455 And might have a higher accretion rate and associated mass-loss rate than other high-state CVs. However, if Model I is closer to the right model for V455 And in outburst, the magnitude of this mass-loss rate, at 40\% of the accretion rate,  represents a big challenge to wind-driving mechanisms in CVs.
Current dynamical simulations of line-driven winds in CVs estimate the mass-loss to accretion rate ratio as $\approx 10^{-3}$ \citep{dre00radiationwind} or even $\lesssim 10^{-5}$ \citep{hig24linedriven} if more realistic physics is accounted for.
Although a clumpy structure in winds can reduce the required net wind mass-loss rate (i.e., Model II; \citealt{ini22j0714}), this poses a different theoretical challenge, because neither the degree of clumping nor the clumping mechanism is well constrained.

The wind acceleration parameters, which in reality should be linked to the wind driving mechanism, strongly impact the optical spectra as seen in the differences between Models A, B, and X (see also \citealt{shl93CVwind}). 
\citetalias{mat15CVdiskwind} (Models B and X) reproduced the RW Tri spectrum by updating the acceleration parameters ($R_V$ and $\alpha$) from Model A. 
In our Models I and II, we instead use the same acceleration parameters as Model A, which was originally designed to reproduce the UV spectra of CVs. 
Thus, while the models adopted for V455~And imply a uniquely high mass-loss rate, the underlying wind kinematics and driving mechanisms may not be particularly different from those in other CVs.
Other spectral synthesis calculations for high-state CVs using \python~ adopted  acceleration parameters $\alpha =$ 1.0 -- 2.2 and $R_V =$ $5\times 10^{10}$ -- $2\times10^{11}$ cm \citep{noe10rwtriuxuma, ini22j0714}.
Moreover, \citet{ini22j0714} adopted a very low filling factor $f_V = 0.008$ to reproduce the strong optical emission lines in a NL ASAS J071404$+$7004.3.
This comparison proposes a relatively low acceleration exponent (and what this implies about wind acceleration physics) may be common to high-state CVs.

In principle, disc winds may also have a significant impact on the evolution of the underlying binary system. 
For example, \citet{can88angularmomentumloss} studied the angular momentum loss through disc winds assuming a magneto-centrifugal wind. 
Although the inferred mass-loss rate of $\simeq$ 10$^{-8}$ M$_\odot$ yr$^{-1}$ in V455~And is 1 -- 2 orders of magnitude higher than those in previously-studied high-state CVs, the ratio of the mass-loss rate over accretion rate is broadly similar once a clumpy wind structure is considered. 
Thus, provided that most of the accretion in DNe takes place during outbursts, any impact of the disc wind on binary evolution is likely to be similar in V455~And and other wind-driving CVs.

\subsection{Can a decline in accretion and mass-loss rates explain the spectral evolution across the outburst?}
\label{subsec:timeevol}

\begin{figure*}
    \centering
    \includegraphics[width=0.49\linewidth]{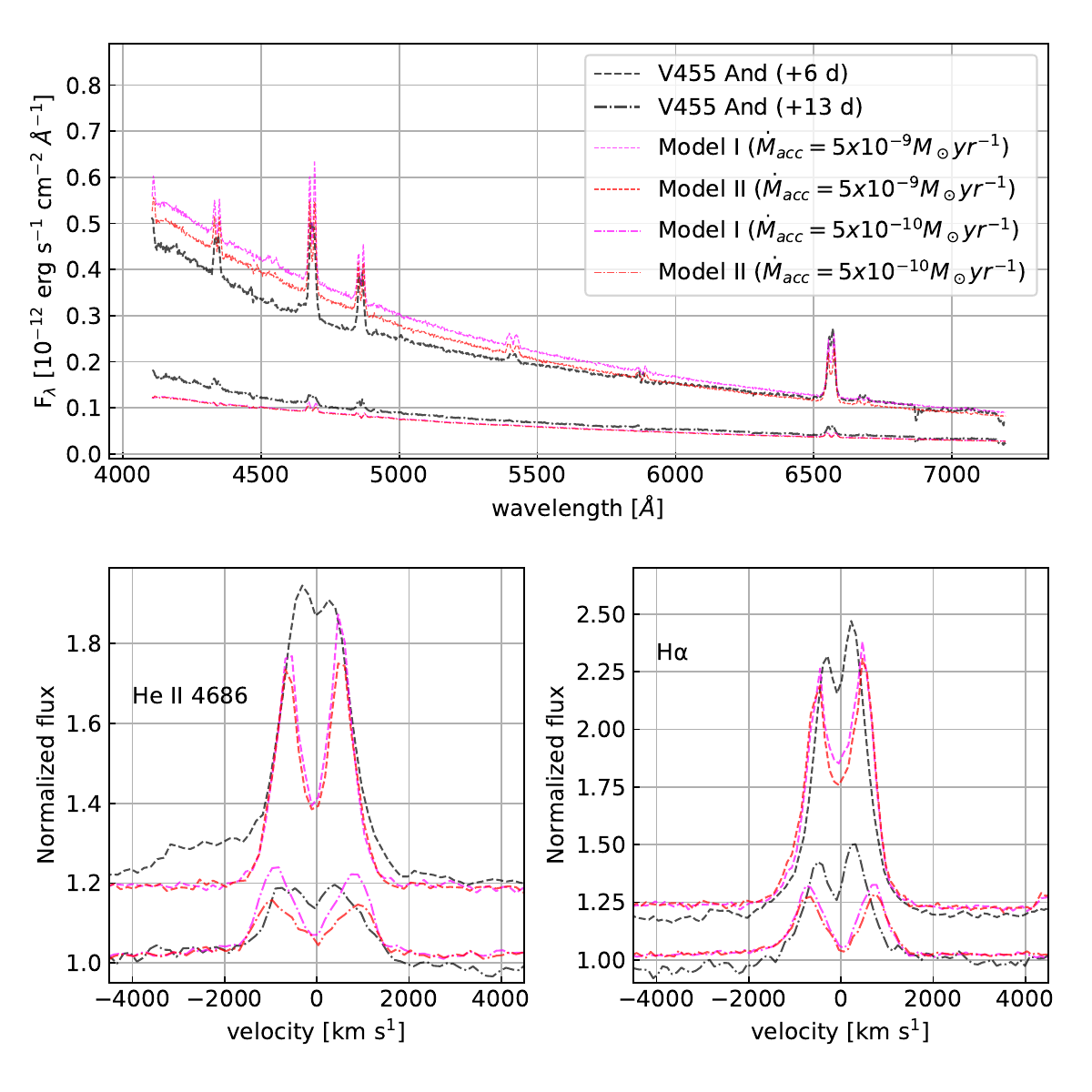}
    \includegraphics[width=0.49\linewidth]{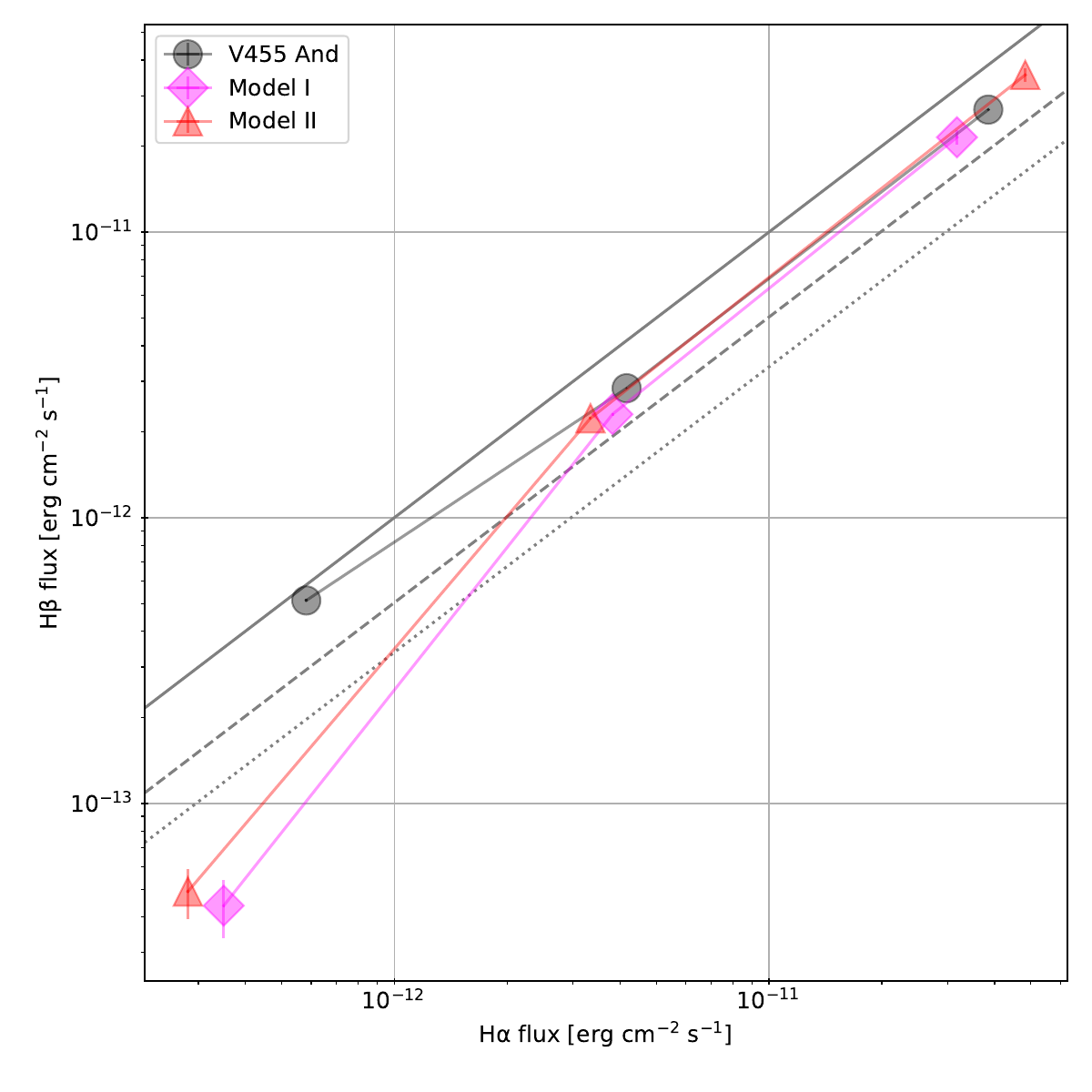}
    \caption{
    Left panel: black dashed and dot-dashed lines represent the observations on BJD 2454356.08 and 2454363.03, respectively.
    Magenta and red lines are the simulated spectra using wind Model I and Model II assuming the disc accretion rate $\dot{M}_{\rm acc}  = 5\times10^{-9}$ (dashed) and $5\times10^{-10}$ M$_\odot$ yr$^{-1}$ (dot-dashed).
    Right panel: the evolution of H$\alpha$ and H$\beta$ line flux of V455~And (black circles), Model I (magenta squares), and Model II (red triangles).
    The thick solid, dashed, and dotted lines represent the flux ratio of 1, 2, and 3 of H$\alpha$ over H$\beta$.
    }
    \label{fig:spec-evol}
\end{figure*}

In terms of studying accretion physics, the biggest advantage of DNe compared to NLs is that DN outbursts are a transient event.
As presented in \citetalias{tam22v455andspec} and  \citet{tov22v455andspec}, the emission line profiles changed greatly throughout the V455~And outburst.
The dashed and dot-dashed lines in figure \ref{fig:v455andspec} show the spectra of V455~And on BJD 2454356.08 (+6 d from the modelled spectrum) and BJD 2454363.03 (+13 d from the modelled spectrum), respectively, taken from \citetalias{tam22v455andspec}.
The enlarged panels show that the single-peaked H$\alpha$ around the outburst maximum changed dramatically into the double-peaked and weaker profiles in later epochs.

To assess whether the decrease of disc accretion and wind mass-loss rates can explain the observed spectral evolution, we conducted additional simulations designed to approximate the V455~And system decaying from outburst maximum.
Although many other parameters (i.e. disc radius, WD temperature) should change throughout the outburst,
for simplicity, we fixed all the binary and wind-related parameters including the $\dot{M}_{\rm wind}/\dot{M}_{\rm acc}$ ratio (40\% and 10\% for Models I and II, respectively) and the volume filling factors, as well as $L_{\rm BL} = L_{\rm disc}$. 
Hence, the only adjustable parameter was the intrinsic disc accretion rate. 
To match the continuum flux level around H$\alpha$, we adopted the disc accretion rates of $\dot{M}_{\rm acc}  = 5\times10^{-9}$ and $5\times10^{-10}$ M$_\odot$ yr$^{-1}$ for the spectra on BJD 2454356.08 and 2454363.03, respectively.
In the left panels of figure \ref{fig:spec-evol}, the magenta (Model I) and red (Model II) lines show the simulated spectra for $\dot{M}_{\rm acc}  = 5\times10^{-9}$ (dashed) and $5\times10^{-10}$ M$_\odot$ yr$^{-1}$ (dot-dashed).
The right panel of figure \ref{fig:spec-evol} presents the evolution of the H$\alpha$ and H$\beta$ line fluxes of V455~And (black circles), Model I (magenta squares), and Model II (red triangles).
These simulated spectra of Models I and II for BJD 2454356.08 reasonably produce the observed Balmer and He~\textsc{ii} line fluxes, EWs, and line FWZIs ($\approx$ 3000 km s$^{-1}$).
The peak separations ($\geq$ 1000 km s$^{-1}$) tend to be larger than the observations, which was also seen in the simulated spectra for those around the outburst maximum.
The simulated spectra for BJD 2454363.03, however, produced both H$\alpha$ and \heiifses~ emission line with the peak separation of $\geq 1500$ km s$^{-1}$, and have a larger discrepancy from the observations in the line fluxes.
Therefore, at this epoch, the optically thin region in the outer disc surface (i.e., disc corona or atmosphere) may become a more dominant source of the emission lines, rather than the disc winds from the inner disc.
\citet{pro99windcomparison} showed the wind mass-loss rate is proportional to the disc accretion rate above a few 10$^{-9}$ M$_\odot$ yr$^{-1}$, while it dramatically drops below this critical accretion rate.
Our calculations also implies that, at an accretion rate $\dot{M}_{\rm acc} \gtrsim 10^{-9}$ M$_\odot$ yr$^{-1}$, 
disc winds with similar kinematics and geometry but with a decreasing accretion rate dominate the spectral evolution in V455~And.

\subsection{Could disc winds with high mass-loss rates be present in other high-state cataclysmic variables?}
\label{subsec:nlimplication}

\begin{figure*}
    \centering
    \includegraphics[width=\linewidth]{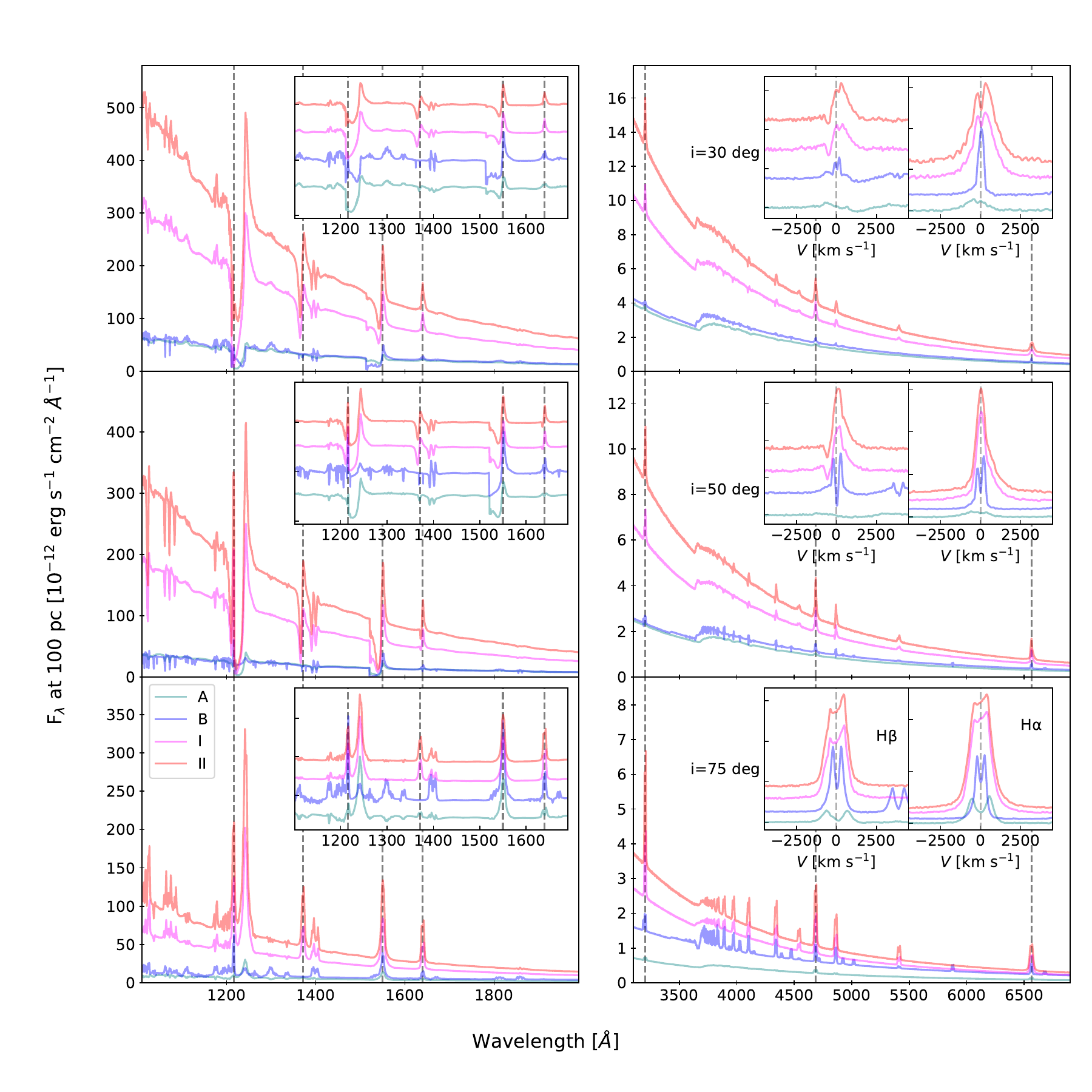}
    \caption{
    The simulated spectra of Model A (green), B (blue), I (magenta), and II (red) in UV (left) and optical (right) wavelength.
    The assumed inclinations are 30$^\circ$, 50$^\circ$, and 75$^\circ$ from upper to lower panels.
    The panels inserted in the UV and optical panels show the normalized UV spectra, and the H$\alpha$ and H$\beta$ profiles, respectively.
    Vertical dashed lines represent the rest wavelengths of Ly$\alpha$, \ovotso, \civoffo, \heiiosfo, \heiittot, \heiifses, and H$\alpha$ from left to right.
    }
    \label{fig:diff-incl}
\end{figure*}

Although our primary focus in this article was on reproducing the V455~And optical spectra at the outburst maximum, \python~ also allows us to investigate the simulated spectrum over a wider wavelength range and at other inclinations.
Figure \ref{fig:diff-incl} shows the calculated spectra of Models A (green), B (blue), I (magenta), and II (red) in the UV (left panels) and optical (right panels) wavelengths.
The panels inserted in UV spectra show the normalized spectrum, and those in the optical spectra show the H$\alpha$ and H$\beta$ profile.

As in the optical, the UV fluxes of Models I and II are brighter than those of Models A and B by a factor of $\sim $3--5.
This is essentially due to the increased disc accretion rate in Models I and II. 
However, the normalized spectrum is less sensitive to the change in the continuum. 
Although Models I and II are not optimized for UV spectra,
for example, the most prominent P-Cygni profiles of \civoffo\ and N~\textsc{v}~$\lambda$1240 have a similar shape to each other between Model A and Models I and II at all the presented inclinations. 
On the other hand, \ovotso\ appears to be very sensitive to the mass-loss rate; Models I and II with $i = 30^\circ$ and $50^\circ$ show clear P-Cygni profiles in O~\textsc{v}~$\lambda$1371, while Models A and B do not.

Moreover, our simulations predict that if Models I and II were observed in lower-inclination systems (e.g., $i$=30$^\circ$ and 50$^\circ$), their H$\alpha$ and H$\beta$ show asymmetric and even P-Cygni profiles.
This is the first time that these optical P-Cygni profiles are produced in simulations of disc wind models with kinematic prescriptions.
Since some NLs are known to show asymmetric and/or P-Cygni profiles in H$\alpha$ with variable absorption depths \citep{kaf04windfromCV, ini22j0714, cun23nloutflow}, this suggests that the characteristic disc wind structure of V455~And is not unique, but may be a common feature, at least in some NLs at some epochs with the highest mass-transfer rates (and hence the highest accretion rates).
One possible explanation for missing these UV and optical trends in high-state CVs in previous literature is that, historically, modelling disc winds in CVs might have been biased to choose lower mass-loss rate models based on poorly constrained distances and inclinations, which is easier to understand both in simulations (e.g. \citealt{pro99windcomparison}) and in real situations.

Some X-ray binaries also show P-Cygni and/or asymmetric line profiles in optical \citep{mun16v404cygwind, mun19j1820wind, mat18v404cygwind} and in the UV \citep{cas22J1858}, which usually implies the existence of comparatively cool, low-ionization material in outflows. { Radiation pressure in all forms (e.g. electron scattering, bound-free, and bound-bound) may contribute to driving these outflows \citep{hig20lmxbhardwind, tom19thermalradwind}, but irradiation-powered thermal expansion \cite[e.g.][]{beg83thermalwind, woo96thermalwind} and/or magnetic/centrifugal forces \cite[e.g.][]{bla82hmflow} is likely to be the dominant driving mechanisms.  \citet{cas22J1858} and soon after \citet{mun22v404cygwind} discuss the co-existence of multi-phase wind structure based on the simultaneous detection of wind-originated features at different wavelengths. Our simulated spectra for V455~And (Figure \ref{fig:diff-incl}) support this idea by showing that UV and optical wind features can be naturally formed in different parts of a single continuous outflow with a stratified thermal and ionization state.}

\section{Summary}
\label{sec:summary}

We have presented spectral synthesis calculations of disc winds in high-state CVs using the Monte Carlo ionization and radiative transfer simulation code \python. Our primary aim is to test whether disc + wind models can account for the observed spectra of the WZ Sge-type DN V455~And near outburst maximum. These spectra displayed a blue continuum punctuated by strong, broad-based, and narrow-peaked emission lines of the Balmer and He~\textsc{ii} series. These line profiles are inconsistent with emission from a rotating accretion disc. If current estimates of the distance and inclination are correct, V445 And is very luminous at outburst maximum for a dwarf nova.  If the continuum emission arises from a steady state disk, an accretion accretion rate of $\dot{M}_{\rm acc} \approx 10^{-7}$ M$_\odot$ yr$^{-1}$ is required. 
Our primary results from modelling these features as arising in a disc wind can be summarized as follows:

\begin{itemize}

\item 
    To explain the spectral features of V455~And, our disc wind models require a wind mass-loss rate $\dot{M}_{\rm wind} \approx 10^{-8}$ M$_\odot$ yr$^{-1}$, and either (1) a high wind mass-loss to disc accretion rate ratio reaching 40\%, and/or (2) a clumpy wind structure with similar volume filling factors as in stellar winds. 
    Such models feature larger line-forming regions with high electron densities and produce relatively strong emission lines, as observed.

\item 
   
    The estimated wind mass-loss rate in V455~And is 1 -- 2 orders of magnitude higher than the previous CV wind models.
    By considering a clumpy wind structure, a wind mass-loss to disc accretion rate ratio can be reduced to 10\%.
    This high mass-loss rate is likely attributed to the high accretion rate of V455~And at its outburst maximum.
    On the other hand, the wind acceleration parameters, which should be associated with the wind driving mechanism, are fairly similar to those used to model disc winds in other CVs.

\item 
    We also calculated wind models assuming the same kinematics but decreased mass-loss and accretion rates. 
    These models reproduce the observed spectra during the decay phase reasonably well, especially in terms of the line fluxes and EWs of the Balmer and He~\textsc{ii} lines.
    Therefore, the disc wind is likely the dominant factor explaining not only the spectra around the outburst maximum but also the overall spectral evolution in V455~And.

\item 
    We also presented simulated spectra for lower-inclination cases.
    Although the UV continuum fluxes in the V455~And wind models are a factor of 3--5 brighter than previous CV wind models, the normalized spectra and P-Cygni profiles in UV wavelengths are much less sensitive to the mass-loss rate than the continuum.
    The Balmer lines in these spectra show asymmetric and even P-Cygni profiles, which have been observed in some (but not all) NLs.
    These points suggest that previous wind models might have been biased towards lower wind mass-loss rates and that winds similar to that in V455~And may be present in some NLs and possibly in some X-ray binaries with clear wind features.

\end{itemize}

\section*{Acknowledgements}

YT acknowledges support from the Japan Society for the Promotion of Science (JSPS) KAKENHI Grant Number 21J22351 and the JSPS Overseas Challenge Program for Young Researchers. 
Partial support for KSL's effort on the project was provided by NASA through grant numbers HST-GO-16489 and HST-GO-16659 and from the Space Telescope Science Institute, which is operated by AURA, Inc., under NASA contract NAS 5-26555.
The authors also thank Stuart A. Sim, Nicolas Scepi, and Austen Wallis for their valuable discussion.
A part of calculations in this work made use of the Iridis 5 Supercomputer at the University of Southampton.


\section*{Data Availability}

The \python~ codes used to perform these simulations are
available from the sites \url{https://github.com/agnwinds/python}.
The data files used to generate the figures presented here are available upon request.



\bibliographystyle{mnras}
\bibliography{cvs}

\newcommand{\noop}[1]{}
\begin{thebibliography}{}
\makeatletter
\relax
\def\mn@urlcharsother{\let\do\@makeother \do\$\do\&\do\#\do\^\do\_\do\%\do\~}
\def\mn@doi{\begingroup\mn@urlcharsother \@ifnextchar [ {\mn@doi@} {\mn@doi@[]}}
\def\mn@doi@[#1]#2{\def\@tempa{#1}\ifx\@tempa\@empty \href {http://dx.doi.org/#2} {doi:#2}\else \href {http://dx.doi.org/#2} {#1}\fi \endgroup}
\def\mn@eprint#1#2{\mn@eprint@#1:#2::\@nil}
\def\mn@eprint@arXiv#1{\href {http://arxiv.org/abs/#1} {{\tt arXiv:#1}}}
\def\mn@eprint@dblp#1{\href {http://dblp.uni-trier.de/rec/bibtex/#1.xml} {dblp:#1}}
\def\mn@eprint@#1:#2:#3:#4\@nil{\def\@tempa {#1}\def\@tempb {#2}\def\@tempc {#3}\ifx \@tempc \@empty \let \@tempc \@tempb \let \@tempb \@tempa \fi \ifx \@tempb \@empty \def\@tempb {arXiv}\fi \@ifundefined {mn@eprint@\@tempb}{\@tempb:\@tempc}{\expandafter \expandafter \csname mn@eprint@\@tempb\endcsname \expandafter{\@tempc}}}

\bibitem[\protect\citeauthoryear{{Araujo-Betancor} et~al.,}{{Araujo-Betancor} et~al.}{2005}]{ara05v455and}
{Araujo-Betancor} S.,  et~al., 2005, \mn@doi [\aap] {10.1051/0004-6361:20041736}, \href {https://ui.adsabs.harvard.edu/abs/2005A&A...430..629A} {430, 629}

\bibitem[\protect\citeauthoryear{{Baba} et~al.,}{{Baba} et~al.}{2002}]{bab02wzsgeletter}
{Baba} H.,  et~al., 2002, \mn@doi [\pasj] {10.1093/pasj/54.1.L7}, \href {https://ui.adsabs.harvard.edu/abs/2002PASJ...54L...7B} {54, L7}

\bibitem[\protect\citeauthoryear{{Badnell}, {Bautista}, {Butler}, {Delahaye}, {Mendoza}, {Palmeri}, {Zeippen}  \& {Seaton}}{{Badnell} et~al.}{2005}]{bad05opacitymodel}
{Badnell} N.~R.,  {Bautista} M.~A.,  {Butler} K.,  {Delahaye} F.,  {Mendoza} C.,  {Palmeri} P.,  {Zeippen} C.~J.,   {Seaton} M.~J.,  2005, \mn@doi [\mnras] {10.1111/j.1365-2966.2005.08991.x}, \href {https://ui.adsabs.harvard.edu/abs/2005MNRAS.360..458B} {360, 458}

\bibitem[\protect\citeauthoryear{{Bailer-Jones}, {Rybizki}, {Fouesneau}, {Demleitner}  \& {Andrae}}{{Bailer-Jones} et~al.}{2021}]{Bai21GaiaEDR3distance}
{Bailer-Jones} C.~A.~L.,  {Rybizki} J.,  {Fouesneau} M.,  {Demleitner} M.,   {Andrae} R.,  2021, \mn@doi [\aj] {10.3847/1538-3881/abd806}, \href {https://ui.adsabs.harvard.edu/abs/2021AJ....161..147B} {161, 147}

\bibitem[\protect\citeauthoryear{{Balman}, {Schlegel}  \& {Godon}}{{Balman} et~al.}{2022}]{bal22bzcamv592casxray}
{Balman} {\c{S}}.,  {Schlegel} E.~M.,   {Godon} P.,  2022, \mn@doi [\apj] {10.3847/1538-4357/ac6616}, \href {https://ui.adsabs.harvard.edu/abs/2022ApJ...932...33B} {932, 33}

\bibitem[\protect\citeauthoryear{{Begelman}, {McKee}  \& {Shields}}{{Begelman} et~al.}{1983}]{beg83thermalwind}
{Begelman} M.~C.,  {McKee} C.~F.,   {Shields} G.~A.,  1983, \mn@doi [\apj] {10.1086/161178}, \href {https://ui.adsabs.harvard.edu/abs/1983ApJ...271...70B} {271, 70}

\bibitem[\protect\citeauthoryear{{Blandford} \& {Payne}}{{Blandford} \& {Payne}}{1982}]{bla82hmflow}
{Blandford} R.~D.,  {Payne} D.~G.,  1982, \mn@doi [\mnras] {10.1093/mnras/199.4.883}, \href {https://ui.adsabs.harvard.edu/abs/1982MNRAS.199..883B} {199, 883}

\bibitem[\protect\citeauthoryear{{Cannizzo} \& {Pudritz}}{{Cannizzo} \& {Pudritz}}{1988}]{can88angularmomentumloss}
{Cannizzo} J.~K.,  {Pudritz} R.~E.,  1988, \mn@doi [\apj] {10.1086/166241}, \href {https://ui.adsabs.harvard.edu/abs/1988ApJ...327..840C} {327, 840}

\bibitem[\protect\citeauthoryear{{Castro Segura} et~al.,}{{Castro Segura} et~al.}{2022}]{cas22J1858}
{Castro Segura} N.,  et~al., 2022, \mn@doi [\nat] {10.1038/s41586-021-04324-2}, \href {https://ui.adsabs.harvard.edu/abs/2022Natur.603...52C} {603, 52}

\bibitem[\protect\citeauthoryear{{Cordova} \& {Mason}}{{Cordova} \& {Mason}}{1982}]{cor82outburstwind}
{Cordova} F.~A.,  {Mason} K.~O.,  1982, \mn@doi [\apj] {10.1086/160291}, \href {https://ui.adsabs.harvard.edu/abs/1982ApJ...260..716C} {260, 716}

\bibitem[\protect\citeauthoryear{{C{\'u}neo} et~al.,}{{C{\'u}neo} et~al.}{2023}]{cun23nloutflow}
{C{\'u}neo} V.~A.,  et~al., 2023, \mn@doi [\aap] {10.1051/0004-6361/202347265}, \href {https://ui.adsabs.harvard.edu/abs/2023A&A...679A..85C} {679, A85}

\bibitem[\protect\citeauthoryear{{Dhillon}, {Smith}  \& {Marsh}}{{Dhillon} et~al.}{2013}]{dhi13swsexenigma}
{Dhillon} V.~S.,  {Smith} D.~A.,   {Marsh} T.~R.,  2013, \mn@doi [\mnras] {10.1093/mnras/sts294}, \href {https://ui.adsabs.harvard.edu/abs/2013MNRAS.428.3559D} {428, 3559}

\bibitem[\protect\citeauthoryear{{Drew}}{{Drew}}{1987}]{dre87CVlineprofile}
{Drew} J.~E.,  1987, \mn@doi [\mnras] {10.1093/mnras/224.3.595}, \href {https://ui.adsabs.harvard.edu/abs/1987MNRAS.224..595D} {224, 595}

\bibitem[\protect\citeauthoryear{{Drew} \& {Proga}}{{Drew} \& {Proga}}{2000}]{dre00radiationwind}
{Drew} J.~E.,  {Proga} D.,  2000, \mn@doi [\nar] {10.1016/S1387-6473(00)00007-5}, \href {https://ui.adsabs.harvard.edu/abs/2000nar..44...21D} {44, 21}

\bibitem[\protect\citeauthoryear{{Dutta}, {Rana}, {Mukai}  \& {de Oliveira}}{{Dutta} et~al.}{2023}]{dut23sscygxray}
{Dutta} A.,  {Rana} V.,  {Mukai} K.,   {de Oliveira} R.~L.,  2023, \mn@doi [\apj] {10.3847/1538-4357/acf838}, \href {https://ui.adsabs.harvard.edu/abs/2023ApJ...957...33D} {957, 33}

\bibitem[\protect\citeauthoryear{{Fontaine}, {Brassard}  \& {Bergeron}}{{Fontaine} et~al.}{2001}]{fon01whitedwarf}
{Fontaine} G.,  {Brassard} P.,   {Bergeron} P.,  2001, \mn@doi [\pasp] {10.1086/319535}, \href {https://ui.adsabs.harvard.edu/abs/2001PASP..113..409F} {113, 409}

\bibitem[\protect\citeauthoryear{{Froning}}{{Froning}}{2005}]{fro05cvoutflow}
{Froning} C.~S.,  2005, in {Hameury} J.~M.,  {Lasota} J.~P.,  eds,  Astronomical Society of the Pacific Conference Series Vol. 330, The Astrophysics of Cataclysmic Variables and Related Objects. p.~81 (\mn@eprint {arXiv} {astro-ph/0410200}), \mn@doi{10.48550/arXiv.astro-ph/0410200}

\bibitem[\protect\citeauthoryear{{Gaia Collaboration} et~al.,}{{Gaia Collaboration} et~al.}{2021}]{gaiaedr3}
{Gaia Collaboration} et~al., 2021, \mn@doi [\aap] {10.1051/0004-6361/202039657}, \href {https://ui.adsabs.harvard.edu/abs/2021A&A...649A...1G} {649, A1}

\bibitem[\protect\citeauthoryear{{Godon}, {Sion}, {Balman}  \& {Blair}}{{Godon} et~al.}{2017}]{god17NLdisk}
{Godon} P.,  {Sion} E.~M.,  {Balman} {\c S}.,   {Blair} W.~P.,  2017, ApJ, 846, 52

\bibitem[\protect\citeauthoryear{{Hamann} \& {Koesterke}}{{Hamann} \& {Koesterke}}{1998}]{ham98stellarwind}
{Hamann} W.~R.,  {Koesterke} L.,  1998, \aap, \href {https://ui.adsabs.harvard.edu/abs/1998A&A...335.1003H} {335, 1003}

\bibitem[\protect\citeauthoryear{{Hamann}, {Koesterke}  \& {Wessolowski}}{{Hamann} et~al.}{1995}]{ham95wrstarclump}
{Hamann} W.~R.,  {Koesterke} L.,   {Wessolowski} U.,  1995, \aap, \href {https://ui.adsabs.harvard.edu/abs/1995A&A...299..151H} {299, 151}

\bibitem[\protect\citeauthoryear{{Hameury}}{{Hameury}}{2020}]{ham20CVreview}
{Hameury} J.~M.,  2020, \mn@doi [Advances in Space Research] {10.1016/j.asr.2019.10.022}, \href {https://ui.adsabs.harvard.edu/abs/2020AdSpR..66.1004H} {66, 1004}

\bibitem[\protect\citeauthoryear{{Heap} et~al.,}{{Heap} et~al.}{1978}]{hea78iuespec}
{Heap} S.~R.,  et~al., 1978, \mn@doi [\nat] {10.1038/275385a0}, \href {https://ui.adsabs.harvard.edu/abs/1978Natur.275..385H} {275, 385}

\bibitem[\protect\citeauthoryear{{Hellier}}{{Hellier}}{1996}]{hel96v1315aql}
{Hellier} C.,  1996, \mn@doi [\apj] {10.1086/178021}, \href {https://ui.adsabs.harvard.edu/abs/1996ApJ...471..949H} {471, 949}

\bibitem[\protect\citeauthoryear{{Hellier}}{{Hellier}}{2001}]{hel01book}
{Hellier} C.,  2001, {Cataclysmic Variable Stars}.
Berlin: Springer

\bibitem[\protect\citeauthoryear{{Higginbottom}, {Knigge}, {Long}, {Sim}  \& {Matthews}}{{Higginbottom} et~al.}{2013}]{hig13balquasar}
{Higginbottom} N.,  {Knigge} C.,  {Long} K.~S.,  {Sim} S.~A.,   {Matthews} J.~H.,  2013, \mn@doi [\mnras] {10.1093/mnras/stt1658}, \href {https://ui.adsabs.harvard.edu/abs/2013MNRAS.436.1390H} {436, 1390}

\bibitem[\protect\citeauthoryear{{Higginbottom}, {Proga}, {Knigge}, {Long}, {Matthews}  \& {Sim}}{{Higginbottom} et~al.}{2014}]{hig14agnwind}
{Higginbottom} N.,  {Proga} D.,  {Knigge} C.,  {Long} K.~S.,  {Matthews} J.~H.,   {Sim} S.~A.,  2014, \mn@doi [\apj] {10.1088/0004-637X/789/1/19}, \href {https://ui.adsabs.harvard.edu/abs/2014ApJ...789...19H} {789, 19}

\bibitem[\protect\citeauthoryear{{Higginbottom}, {Knigge}, {Sim}, {Long}, {Matthews}, {Hewitt}, {Parkinson}  \& {Mangham}}{{Higginbottom} et~al.}{2020}]{hig20lmxbhardwind}
{Higginbottom} N.,  {Knigge} C.,  {Sim} S.~A.,  {Long} K.~S.,  {Matthews} J.~H.,  {Hewitt} H.~A.,  {Parkinson} E.~J.,   {Mangham} S.~W.,  2020, \mn@doi [\mnras] {10.1093/mnras/staa209}, \href {https://ui.adsabs.harvard.edu/abs/2020MNRAS.492.5271H} {492, 5271}

\bibitem[\protect\citeauthoryear{{Higginbottom}, {Scepi}, {Knigge}, {Long}, {Matthews}  \& {Sim}}{{Higginbottom} et~al.}{2024}]{hig24linedriven}
{Higginbottom} N.,  {Scepi} N.,  {Knigge} C.,  {Long} K.~S.,  {Matthews} J.~H.,   {Sim} S.~A.,  2024, \mn@doi [\mnras] {10.1093/mnras/stad3830}, \href {https://ui.adsabs.harvard.edu/abs/2024MNRAS.527.9236H} {527, 9236}

\bibitem[\protect\citeauthoryear{{Hillier} \& {Miller}}{{Hillier} \& {Miller}}{1999}]{hil99hd165763}
{Hillier} D.~J.,  {Miller} D.~L.,  1999, \mn@doi [\apj] {10.1086/307339}, \href {https://ui.adsabs.harvard.edu/abs/1999ApJ...519..354H} {519, 354}

\bibitem[\protect\citeauthoryear{{Hoard}, {Szkody}, {Froning}, {Long}  \& {Knigge}}{{Hoard} et~al.}{2003}]{hoa03dwuma}
{Hoard} D.~W.,  {Szkody} P.,  {Froning} C.~S.,  {Long} K.~S.,   {Knigge} C.,  2003, \mn@doi [\aj] {10.1086/378605}, \href {https://ui.adsabs.harvard.edu/abs/2003AJ....126.2473H} {126, 2473}

\bibitem[\protect\citeauthoryear{{Hoare}}{{Hoare}}{1994}]{hoa94CVHeII}
{Hoare} M.~G.,  1994, \mn@doi [\mnras] {10.1093/mnras/267.1.153}, \href {https://ui.adsabs.harvard.edu/abs/1994MNRAS.267..153H} {267, 153}

\bibitem[\protect\citeauthoryear{{Honeycutt}, {Schlegel}  \& {Kaitchuck}}{{Honeycutt} et~al.}{1986}]{hon86swsex}
{Honeycutt} R.~K.,  {Schlegel} E.~M.,   {Kaitchuck} R.~H.,  1986, \mn@doi [\apj] {10.1086/163997}, \href {https://ui.adsabs.harvard.edu/abs/1986ApJ...302..388H} {302, 388}

\bibitem[\protect\citeauthoryear{{Hubeny} \& {Lanz}}{{Hubeny} \& {Lanz}}{1995}]{hub95nonlte}
{Hubeny} I.,  {Lanz} T.,  1995, \mn@doi [\apj] {10.1086/175226}, \href {https://ui.adsabs.harvard.edu/abs/1995ApJ...439..875H} {439, 875}

\bibitem[\protect\citeauthoryear{{Hubeny} \& {Long}}{{Hubeny} \& {Long}}{2021}]{hub21ixvel}
{Hubeny} I.,  {Long} K.~S.,  2021, \mn@doi [\mnras] {10.1093/mnras/stab830}, \href {https://ui.adsabs.harvard.edu/abs/2021MNRAS.503.5534H} {503, 5534}

\bibitem[\protect\citeauthoryear{{Inight} et~al.,}{{Inight} et~al.}{2022}]{ini22j0714}
{Inight} K.,  et~al., 2022, \mn@doi [\mnras] {10.1093/mnras/stab3662}, \href {https://ui.adsabs.harvard.edu/abs/2022MNRAS.510.3605I} {510, 3605}

\bibitem[\protect\citeauthoryear{{Kafka} \& {Honeycutt}}{{Kafka} \& {Honeycutt}}{2004}]{kaf04windfromCV}
{Kafka} S.,  {Honeycutt} R.~K.,  2004, \mn@doi [\aj] {10.1086/424618}, \href {https://ui.adsabs.harvard.edu/abs/2004AJ....128.2420K} {128, 2420}

\bibitem[\protect\citeauthoryear{{Kato}}{{Kato}}{2015}]{kat15wzsge}
{Kato} T.,  2015, \mn@doi [\pasj] {10.1093/pasj/psv077}, \href {https://ui.adsabs.harvard.edu/abs/2015PASJ...67..108K} {67, 108}

\bibitem[\protect\citeauthoryear{{Kato} \& {Osaki}}{{Kato} \& {Osaki}}{2013}]{kat13qfromstageA}
{Kato} T.,  {Osaki} Y.,  2013, \mn@doi [\pasj] {10.1093/pasj/65.6.115}, \href {https://ui.adsabs.harvard.edu/abs/2013PASJ...65..115K} {65, 115}

\bibitem[\protect\citeauthoryear{{Kato} et~al.,}{{Kato} et~al.}{2009}]{Pdot}
{Kato} T.,  et~al., 2009, \mn@doi [\pasj] {10.1093/pasj/61.sp2.S395}, \href {https://ui.adsabs.harvard.edu/abs/2009PASJ...61S.395K} {61, S395}

\bibitem[\protect\citeauthoryear{{Kimura}}{{Kimura}}{2020}]{kim20thesis}
{Kimura} M.,  2020, {Observational and Theoretical Studies on Dwarf-nova Outbursts}, 1 edn.
Springer Theses, Springer Singapore, \mn@doi{10.1007/978-981-15-8912-6}

\bibitem[\protect\citeauthoryear{{Kimura}, {Kashiyama}, {Shigeyama}, {Tampo}, {Yamada}  \& {Enoto}}{{Kimura} et~al.}{2023}]{kim23j0302}
{Kimura} M.,  {Kashiyama} K.,  {Shigeyama} T.,  {Tampo} Y.,  {Yamada} S.,   {Enoto} T.,  2023, \mn@doi [\apj] {10.3847/1538-4357/acd933}, \href {https://ui.adsabs.harvard.edu/abs/2023ApJ...951..124K} {951, 124}

\bibitem[\protect\citeauthoryear{{King}, {Rolfe}  \& {Schenker}}{{King} et~al.}{2003}]{kin03dntypeiasn}
{King} A.~R.,  {Rolfe} D.~J.,   {Schenker} K.,  2003, \mn@doi [\mnras] {10.1046/j.1365-8711.2003.06639.x}, \href {https://ui.adsabs.harvard.edu/abs/2003MNRAS.341L..35K} {341, L35}

\bibitem[\protect\citeauthoryear{{Knigge} \& {Drew}}{{Knigge} \& {Drew}}{1997}]{kni97uxumawindmapping}
{Knigge} C.,  {Drew} J.~E.,  1997, \mn@doi [\apj] {10.1086/304519}, \href {https://ui.adsabs.harvard.edu/abs/1997ApJ...486..445K} {486, 445}

\bibitem[\protect\citeauthoryear{{Knigge}, {Woods}  \& {Drew}}{{Knigge} et~al.}{1995}]{kni95cvwind}
{Knigge} C.,  {Woods} J.~A.,   {Drew} J.~E.,  1995, \mn@doi [\mnras] {10.1093/mnras/273.2.225}, \href {https://ui.adsabs.harvard.edu/abs/1995MNRAS.273..225K} {273, 225}

\bibitem[\protect\citeauthoryear{{Knigge}, {Long}, {Blair}  \& {Wade}}{{Knigge} et~al.}{1997}]{kni97zcam}
{Knigge} C.,  {Long} K.~S.,  {Blair} W.~P.,   {Wade} R.~A.,  1997, \mn@doi [\apj] {10.1086/303607}, \href {https://ui.adsabs.harvard.edu/abs/1997ApJ...476..291K} {476, 291}

\bibitem[\protect\citeauthoryear{{Koljonen}, {Long}, {Matthews}  \& {Knigge}}{{Koljonen} et~al.}{2023}]{kol23j1820wind}
{Koljonen} K.~I.~I.,  {Long} K.~S.,  {Matthews} J.~H.,   {Knigge} C.,  2023, \mn@doi [\mnras] {10.1093/mnras/stad809}, \href {https://ui.adsabs.harvard.edu/abs/2023MNRAS.521.4190K} {521, 4190}

\bibitem[\protect\citeauthoryear{{Krautter}, {Klare}, {Wolf}, {Duerbeck}, {Rahe}, {Vogt}  \& {Wargau}}{{Krautter} et~al.}{1981}]{kra81CVIUE}
{Krautter} J.,  {Klare} G.,  {Wolf} B.,  {Duerbeck} H.~W.,  {Rahe} J.,  {Vogt} N.,   {Wargau} W.,  1981, \aap, \href {https://ui.adsabs.harvard.edu/abs/1981A&A...102..337K} {102, 337}

\bibitem[\protect\citeauthoryear{{Kurucz}}{{Kurucz}}{1991}]{kur91stellaratmospheres}
{Kurucz} R.~L.,  1991, in {Crivellari} L.,  {Hubeny} I.,   {Hummer} D.~G.,  eds,  NATO Advanced Study Institute (ASI) Series C Vol. 341, Stellar Atmospheres - Beyond Classical Models. p.~441, \mn@doi{10.1007/978-94-011-3554-2_42}

\bibitem[\protect\citeauthoryear{{Kuulkers}, {Knigge}, {Steeghs}, {Wheatley}  \& {Long}}{{Kuulkers} et~al.}{2002}]{kuu02wzsge}
{Kuulkers} E.,  {Knigge} C.,  {Steeghs} D.,  {Wheatley} P.~J.,   {Long} K.~S.,  2002, in {G{\"a}nsicke} B.~T.,  {Beuermann} K.,   {Reinsch} K.,  eds,  Astronomical Society of the Pacific Conference Series Vol. 261, The Physics of Cataclysmic Variables and Related Objects. ASP Conference Proceedings, p.~443 (\mn@eprint {arXiv} {astro-ph/0110064}), \mn@doi{10.48550/arXiv.astro-ph/0110064}

\bibitem[\protect\citeauthoryear{{La Dous}}{{La Dous}}{1991}]{lad91CVIUE}
{La Dous} C.,  1991, \aap, \href {https://ui.adsabs.harvard.edu/abs/1991A&A...252..100L} {252, 100}

\bibitem[\protect\citeauthoryear{{Lasota}}{{Lasota}}{2001}]{las01DIDNXT}
{Lasota} J.-P.,  2001, \mn@doi [\nar] {10.1016/S1387-6473(01)00112-9}, \href {https://ui.adsabs.harvard.edu/abs/2001nar..45..449L} {45, 449}

\bibitem[\protect\citeauthoryear{{Lin} \& {Papaloizou}}{{Lin} \& {Papaloizou}}{1979}]{lin79lowqdisk}
{Lin} D.~N.~C.,  {Papaloizou} J.,  1979, \mn@doi [\mnras] {10.1093/mnras/186.4.799}, \href {https://ui.adsabs.harvard.edu/abs/1979MNRAS.186..799L} {186, 799}

\bibitem[\protect\citeauthoryear{{Long} \& {Knigge}}{{Long} \& {Knigge}}{2002}]{lon02python}
{Long} K.~S.,  {Knigge} C.,  2002, \mn@doi [\apj] {10.1086/342879}, \href {https://ui.adsabs.harvard.edu/abs/2002ApJ...579..725L} {579, 725}

\bibitem[\protect\citeauthoryear{{Long}, {Wade}, {Blair}, {Davidsen}  \& {Hubeny}}{{Long} et~al.}{1994}]{lon94ixvel}
{Long} K.~S.,  {Wade} R.~A.,  {Blair} W.~P.,  {Davidsen} A.~F.,   {Hubeny} I.,  1994, \mn@doi [\apj] {10.1086/174107}, \href {https://ui.adsabs.harvard.edu/abs/1994ApJ...426..704L} {426, 704}

\bibitem[\protect\citeauthoryear{{Long}, {Mauche}, {Raymond}, {Szkody}  \& {Mattei}}{{Long} et~al.}{1996}]{lon96ueveugem}
{Long} K.~S.,  {Mauche} C.~W.,  {Raymond} J.~C.,  {Szkody} P.,   {Mattei} J.~A.,  1996, \mn@doi [\apj] {10.1086/177832}, \href {https://ui.adsabs.harvard.edu/abs/1996ApJ...469..841L} {469, 841}

\bibitem[\protect\citeauthoryear{{Lucy}}{{Lucy}}{2002}]{luc02macroatomI}
{Lucy} L.~B.,  2002, \mn@doi [\aap] {10.1051/0004-6361:20011756}, \href {https://ui.adsabs.harvard.edu/abs/2002A&A...384..725L} {384, 725}

\bibitem[\protect\citeauthoryear{{Lucy}}{{Lucy}}{2003}]{luc03macroatomII}
{Lucy} L.~B.,  2003, \mn@doi [\aap] {10.1051/0004-6361:20030357}, \href {https://ui.adsabs.harvard.edu/abs/2003A&A...403..261L} {403, 261}

\bibitem[\protect\citeauthoryear{{Mart{\'\i}nez-Pais}, {Rodr{\'\i}guez-Gil}  \& {Casares}}{{Mart{\'\i}nez-Pais} et~al.}{1999}]{mar99lspeg}
{Mart{\'\i}nez-Pais} I.~G.,  {Rodr{\'\i}guez-Gil} P.,   {Casares} J.,  1999, \mn@doi [\mnras] {10.1046/j.1365-8711.1999.02483.x}, \href {https://ui.adsabs.harvard.edu/abs/1999MNRAS.305..661M} {305, 661}

\bibitem[\protect\citeauthoryear{{Mata S{\'a}nchez} et~al.,}{{Mata S{\'a}nchez} et~al.}{2018}]{mat18v404cygwind}
{Mata S{\'a}nchez} D.,  et~al., 2018, \mn@doi [\mnras] {10.1093/mnras/sty2402}, \href {https://ui.adsabs.harvard.edu/abs/2018MNRAS.481.2646M} {481, 2646}

\bibitem[\protect\citeauthoryear{{Matsui} et~al.,}{{Matsui} et~al.}{2009}]{mat09v455and}
{Matsui} R.,  et~al., 2009, \mn@doi [\pasj] {10.1093/pasj/61.5.1081}, \href {https://ui.adsabs.harvard.edu/abs/2009PASJ...61.1081M} {61, 1081}

\bibitem[\protect\citeauthoryear{{Matthews}, {Knigge}, {Long}, {Sim}  \& {Higginbottom}}{{Matthews} et~al.}{2015}]{mat15CVdiskwind}
{Matthews} J.~H.,  {Knigge} C.,  {Long} K.~S.,  {Sim} S.~A.,   {Higginbottom} N.,  2015, \mn@doi [\mnras] {10.1093/mnras/stv867}, \href {https://ui.adsabs.harvard.edu/abs/2015MNRAS.450.3331M} {450, 3331}

\bibitem[\protect\citeauthoryear{{Matthews}, {Knigge}, {Long}, {Sim}, {Higginbottom}  \& {Mangham}}{{Matthews} et~al.}{2016}]{mat16quasarclumpy}
{Matthews} J.~H.,  {Knigge} C.,  {Long} K.~S.,  {Sim} S.~A.,  {Higginbottom} N.,   {Mangham} S.~W.,  2016, \mn@doi [\mnras] {10.1093/mnras/stw323}, \href {https://ui.adsabs.harvard.edu/abs/2016MNRAS.458..293M} {458, 293}

\bibitem[\protect\citeauthoryear{{Matthews}, {Knigge}  \& {Long}}{{Matthews} et~al.}{2017}]{mat17quasarspec}
{Matthews} J.~H.,  {Knigge} C.,   {Long} K.~S.,  2017, \mn@doi [\mnras] {10.1093/mnras/stx231}, \href {https://ui.adsabs.harvard.edu/abs/2017MNRAS.467.2571M} {467, 2571}

\bibitem[\protect\citeauthoryear{{Matthews}, {Knigge}, {Higginbottom}, {Long}, {Sim}, {Mangham}, {Parkinson}  \& {Hewitt}}{{Matthews} et~al.}{2020}]{mat20agnwind}
{Matthews} J.~H.,  {Knigge} C.,  {Higginbottom} N.,  {Long} K.~S.,  {Sim} S.~A.,  {Mangham} S.~W.,  {Parkinson} E.~J.,   {Hewitt} H.~A.,  2020, \mn@doi [\mnras] {10.1093/mnras/staa136}, \href {https://ui.adsabs.harvard.edu/abs/2020MNRAS.492.5540M} {492, 5540}

\bibitem[\protect\citeauthoryear{{Mauche}}{{Mauche}}{2002}]{mau02CVinEUVE}
{Mauche} C.~W.,  2002, in {Howell} S.~B.,  {Dupuis} J.,  {Golombek} D.,  {Walter} F.~M.,   {Cullison} J.,  eds,  Astronomical Society of the Pacific Conference Series Vol. 264, Continuing the Challenge of EUV Astronomy: Current Analysis and Prospects for the Future. p.~75 (\mn@eprint {arXiv} {astro-ph/0109133}), \mn@doi{10.48550/arXiv.astro-ph/0109133}

\bibitem[\protect\citeauthoryear{{Mauche} \& {Raymond}}{{Mauche} \& {Raymond}}{1987}]{mau87hlcmaIUE}
{Mauche} C.~W.,  {Raymond} J.~C.,  1987, \mn@doi [\apj] {10.1086/165865}, \href {https://ui.adsabs.harvard.edu/abs/1987ApJ...323..690M} {323, 690}

\bibitem[\protect\citeauthoryear{{Mauche} \& {Raymond}}{{Mauche} \& {Raymond}}{1997}]{mau97cvwindreview}
{Mauche} C.~W.,  {Raymond} J.~C.,  1997, in Cosmic Winds and the Heliosphere. p.~111 (\mn@eprint {arXiv} {astro-ph/9702219}), \mn@doi{10.48550/arXiv.astro-ph/9702219}

\bibitem[\protect\citeauthoryear{{Mineshige} \& {Osaki}}{{Mineshige} \& {Osaki}}{1985}]{min85DNDI}
{Mineshige} S.,  {Osaki} Y.,  1985, \pasj, \href {https://ui.adsabs.harvard.edu/abs/1985PASJ...37....1M} {37, 1}

\bibitem[\protect\citeauthoryear{{Mu{\~n}oz-Darias} \& {Ponti}}{{Mu{\~n}oz-Darias} \& {Ponti}}{2022}]{mun22v404cygwind}
{Mu{\~n}oz-Darias} T.,  {Ponti} G.,  2022, \mn@doi [\aap] {10.1051/0004-6361/202243769}, \href {https://ui.adsabs.harvard.edu/abs/2022A&A...664A.104M} {664, A104}

\bibitem[\protect\citeauthoryear{{Mu{\~n}oz-Darias} et~al.,}{{Mu{\~n}oz-Darias} et~al.}{2016}]{mun16v404cygwind}
{Mu{\~n}oz-Darias} T.,  et~al., 2016, \mn@doi [\nat] {10.1038/nature17446}, \href {https://ui.adsabs.harvard.edu/abs/2016Natur.534...75M} {534, 75}

\bibitem[\protect\citeauthoryear{{Mu{\~n}oz-Darias} et~al.,}{{Mu{\~n}oz-Darias} et~al.}{2019}]{mun19j1820wind}
{Mu{\~n}oz-Darias} T.,  et~al., 2019, \mn@doi [\apjl] {10.3847/2041-8213/ab2768}, \href {https://ui.adsabs.harvard.edu/abs/2019ApJ...879L...4M} {879, L4}

\bibitem[\protect\citeauthoryear{{Mukai}}{{Mukai}}{2017}]{muk17xrayawd}
{Mukai} K.,  2017, \mn@doi [\pasp] {10.1088/1538-3873/aa6736}, \href {https://ui.adsabs.harvard.edu/abs/2017PASP..129f2001M} {129, 062001}

\bibitem[\protect\citeauthoryear{{Murray} \& {Chiang}}{{Murray} \& {Chiang}}{1996}]{mur96diskwind}
{Murray} N.,  {Chiang} J.,  1996, \mn@doi [\nat] {10.1038/382789a0}, \href {https://ui.adsabs.harvard.edu/abs/1996Natur.382..789M} {382, 789}

\bibitem[\protect\citeauthoryear{{Murray} \& {Chiang}}{{Murray} \& {Chiang}}{1997}]{mur97diskwind}
{Murray} N.,  {Chiang} J.,  1997, \mn@doi [\apj] {10.1086/303443}, \href {https://ui.adsabs.harvard.edu/abs/1997ApJ...474...91M} {474, 91}

\bibitem[\protect\citeauthoryear{{Noebauer}, {Long}, {Sim}  \& {Knigge}}{{Noebauer} et~al.}{2010}]{noe10rwtriuxuma}
{Noebauer} U.~M.,  {Long} K.~S.,  {Sim} S.~A.,   {Knigge} C.,  2010, \mn@doi [\apj] {10.1088/0004-637X/719/2/1932}, \href {https://ui.adsabs.harvard.edu/abs/2010ApJ...719.1932N} {719, 1932}

\bibitem[\protect\citeauthoryear{{Nogami} \& {Iijima}}{{Nogami} \& {Iijima}}{2004}]{nog04wzsgespec}
{Nogami} D.,  {Iijima} T.,  2004, \mn@doi [\pasj] {10.1093/pasj/56.sp1.S163}, \href {https://ui.adsabs.harvard.edu/abs/2004PASJ...56S.163N} {56, S163}

\bibitem[\protect\citeauthoryear{{Osaki}}{{Osaki}}{1995}]{osa95wzsge}
{Osaki} Y.,  1995, \pasj, \href {https://ui.adsabs.harvard.edu/abs/1995PASJ...47...47O} {47, 47}

\bibitem[\protect\citeauthoryear{{Osaki}}{{Osaki}}{1996}]{osa96review}
{Osaki} Y.,  1996, \mn@doi [\pasp] {10.1086/133689}, \href {https://ui.adsabs.harvard.edu/abs/1996PASP..108...39O} {108, 39}

\bibitem[\protect\citeauthoryear{{Osaki} \& {Meyer}}{{Osaki} \& {Meyer}}{2002}]{osa02wzsgehump}
{Osaki} Y.,  {Meyer} F.,  2002, \mn@doi [\aap] {10.1051/0004-6361:20011744}, \href {https://ui.adsabs.harvard.edu/abs/2002A&A...383..574O} {383, 574}

\bibitem[\protect\citeauthoryear{{Osterbrock} \& {Ferland}}{{Osterbrock} \& {Ferland}}{2006}]{ost06book}
{Osterbrock} D.~E.,  {Ferland} G.~J.,  2006, {Astrophysics of gaseous nebulae and active galactic nuclei}.
University Science Books

\bibitem[\protect\citeauthoryear{{Paczynski}}{{Paczynski}}{1977}]{pac77ADmodel}
{Paczynski} B.,  1977, \mn@doi [\apj] {10.1086/155526}, \href {https://ui.adsabs.harvard.edu/abs/1977ApJ...216..822P} {216, 822}

\bibitem[\protect\citeauthoryear{{Pala} et~al.,}{{Pala} et~al.}{2022}]{pal22WDinCVs}
{Pala} A.~F.,  et~al., 2022, \mn@doi [\mnras] {10.1093/mnras/stab3449}, \href {https://ui.adsabs.harvard.edu/abs/2022MNRAS.510.6110P} {510, 6110}

\bibitem[\protect\citeauthoryear{{Parkinson}, {Knigge}, {Long}, {Matthews}, {Higginbottom}, {Sim}  \& {Hewitt}}{{Parkinson} et~al.}{2020}]{par20tdewind}
{Parkinson} E.~J.,  {Knigge} C.,  {Long} K.~S.,  {Matthews} J.~H.,  {Higginbottom} N.,  {Sim} S.~A.,   {Hewitt} H.~A.,  2020, \mn@doi [\mnras] {10.1093/mnras/staa1060}, \href {https://ui.adsabs.harvard.edu/abs/2020MNRAS.494.4914P} {494, 4914}

\bibitem[\protect\citeauthoryear{{Parkinson}, {Knigge}, {Matthews}, {Long}, {Higginbottom}, {Sim}  \& {Mangham}}{{Parkinson} et~al.}{2022}]{par22tdeopticalwind}
{Parkinson} E.~J.,  {Knigge} C.,  {Matthews} J.~H.,  {Long} K.~S.,  {Higginbottom} N.,  {Sim} S.~A.,   {Mangham} S.~W.,  2022, \mn@doi [\mnras] {10.1093/mnras/stac027}, \href {https://ui.adsabs.harvard.edu/abs/2022MNRAS.510.5426P} {510, 5426}

\bibitem[\protect\citeauthoryear{{Pereyra}, {Kallman}  \& {Blondin}}{{Pereyra} et~al.}{1997}]{per97linewind}
{Pereyra} N.~A.,  {Kallman} T.~R.,   {Blondin} J.~M.,  1997, \mn@doi [\apj] {10.1086/303671}, \href {https://ui.adsabs.harvard.edu/abs/1997ApJ...477..368P} {477, 368}

\bibitem[\protect\citeauthoryear{{Piro}, {Arras}  \& {Bildsten}}{{Piro} et~al.}{2005}]{pir05wdcooling}
{Piro} A.~L.,  {Arras} P.,   {Bildsten} L.,  2005, \mn@doi [\apj] {10.1086/430588}, \href {https://ui.adsabs.harvard.edu/abs/2005ApJ...628..401P} {628, 401}

\bibitem[\protect\citeauthoryear{{Polidan}, {Mauche}  \& {Wade}}{{Polidan} et~al.}{1990}]{pol90cveuv}
{Polidan} R.~S.,  {Mauche} C.~W.,   {Wade} R.~A.,  1990, \mn@doi [\apj] {10.1086/168831}, \href {https://ui.adsabs.harvard.edu/abs/1990ApJ...356..211P} {356, 211}

\bibitem[\protect\citeauthoryear{{Pringle}}{{Pringle}}{1977}]{pri77softXinDN}
{Pringle} J.~E.,  1977, \mn@doi [\mnras] {10.1093/mnras/178.2.195}, \href {https://ui.adsabs.harvard.edu/abs/1977MNRAS.178..195P} {178, 195}

\bibitem[\protect\citeauthoryear{{Prinja} \& {Rosen}}{{Prinja} \& {Rosen}}{1995}]{pri95CVwind}
{Prinja} R.~K.,  {Rosen} R.,  1995, \mn@doi [\mnras] {10.1093/mnras/273.2.461}, \href {https://ui.adsabs.harvard.edu/abs/1995MNRAS.273..461P} {273, 461}

\bibitem[\protect\citeauthoryear{{Prinja}, {Ringwald}, {Wade}  \& {Knigge}}{{Prinja} et~al.}{2000}]{pri00bzcamwindHST}
{Prinja} R.~K.,  {Ringwald} F.~A.,  {Wade} R.~A.,   {Knigge} C.,  2000, \mn@doi [\mnras] {10.1046/j.1365-8711.2000.03111.x}, \href {https://ui.adsabs.harvard.edu/abs/2000MNRAS.312..316P} {312, 316}

\bibitem[\protect\citeauthoryear{{Proga}}{{Proga}}{1999}]{pro99windcomparison}
{Proga} D.,  1999, \mn@doi [\mnras] {10.1046/j.1365-8711.1999.02408.x}, \href {https://ui.adsabs.harvard.edu/abs/1999MNRAS.304..938P} {304, 938}

\bibitem[\protect\citeauthoryear{{Proga}}{{Proga}}{2005}]{pro05cvoutflow}
{Proga} D.,  2005, in {Hameury} J.~M.,  {Lasota} J.~P.,  eds,  Astronomical Society of the Pacific Conference Series Vol. 330, The Astrophysics of Cataclysmic Variables and Related Objects. p.~103 (\mn@eprint {arXiv} {astro-ph/0411200}), \mn@doi{10.48550/arXiv.astro-ph/0411200}

\bibitem[\protect\citeauthoryear{{Proga}, {Stone}  \& {Drew}}{{Proga} et~al.}{1998}]{pro98radiationwind}
{Proga} D.,  {Stone} J.~M.,   {Drew} J.~E.,  1998, \mn@doi [\mnras] {10.1046/j.1365-8711.1998.01337.x}, \href {https://ui.adsabs.harvard.edu/abs/1998MNRAS.295..595P} {295, 595}

\bibitem[\protect\citeauthoryear{{Proga}, {Stone}  \& {Drew}}{{Proga} et~al.}{1999}]{pro99radiationwind}
{Proga} D.,  {Stone} J.~M.,   {Drew} J.~E.,  1999, \mn@doi [\mnras] {10.1046/j.1365-8711.1999.02935.x}, \href {https://ui.adsabs.harvard.edu/abs/1999MNRAS.310..476P} {310, 476}

\bibitem[\protect\citeauthoryear{{Pudritz}, {Ouyed}, {Fendt}  \& {Brandenburg}}{{Pudritz} et~al.}{2007}]{pud07outflows}
{Pudritz} R.~E.,  {Ouyed} R.,  {Fendt} C.,   {Brandenburg} A.,  2007, in {Reipurth} B.,  {Jewitt} D.,   {Keil} K.,  eds, Protostars and Planets V. {Tucson: University of Arizona Press}, p.~277 (\mn@eprint {arXiv} {astro-ph/0603592}), \mn@doi{10.48550/arXiv.astro-ph/0603592}

\bibitem[\protect\citeauthoryear{{Ribeiro} \& {Diaz}}{{Ribeiro} \& {Diaz}}{2008}]{rib08CVflickering}
{Ribeiro} F. M.~A.,  {Diaz} M.~P.,  2008, \mn@doi [\pasj] {10.1093/pasj/60.2.327}, \href {https://ui.adsabs.harvard.edu/abs/2008PASJ...60..327R} {60, 327}

\bibitem[\protect\citeauthoryear{{Rodr{\'\i}guez-Gil}, {Mart{\'\i}nez-Pais}, {Casares}, {Villada}  \& {van Zyl}}{{Rodr{\'\i}guez-Gil} et~al.}{2001}]{rod01v348pup}
{Rodr{\'\i}guez-Gil} P.,  {Mart{\'\i}nez-Pais} I.~G.,  {Casares} J.,  {Villada} M.,   {van Zyl} L.,  2001, \mn@doi [\mnras] {10.1046/j.1365-8711.2001.04965.x}, \href {https://ui.adsabs.harvard.edu/abs/2001MNRAS.328..903R} {328, 903}

\bibitem[\protect\citeauthoryear{{Saitou}, {Tsujimoto}, {Ebisawa}  \& {Ishida}}{{Saitou} et~al.}{2012}]{sai12zcam}
{Saitou} K.,  {Tsujimoto} M.,  {Ebisawa} K.,   {Ishida} M.,  2012, \mn@doi [\pasj] {10.1093/pasj/64.4.88}, \href {https://ui.adsabs.harvard.edu/abs/2012PASJ...64...88S} {64, 88}

\bibitem[\protect\citeauthoryear{{Schmidtobreick}}{{Schmidtobreick}}{2015}]{sch15swsex}
{Schmidtobreick} L.,  2015, in The Golden Age of Cataclysmic Variables and Related Objects - III (Golden2015). Proceedings of Science, p.~34, \mn@doi{10.22323/1.255.0034}

\bibitem[\protect\citeauthoryear{{Senziani}, {Skinner}  \& {Jean}}{{Senziani} et~al.}{2008}]{sen08v455andhardX}
{Senziani} F.,  {Skinner} G.,   {Jean} P.,  2008, The Astronomer's Telegram, \href {https://ui.adsabs.harvard.edu/abs/2008ATel.1372....1S} {1372, 1}

\bibitem[\protect\citeauthoryear{{Shakura} \& {Sunyaev}}{{Shakura} \& {Sunyaev}}{1973}]{sha73alphadisk}
{Shakura} N.~I.,  {Sunyaev} R.~A.,  1973, \aap, \href {https://ui.adsabs.harvard.edu/abs/1973A&A....24..337S} {500, 33}

\bibitem[\protect\citeauthoryear{{Shlosman} \& {Vitello}}{{Shlosman} \& {Vitello}}{1993}]{shl93CVwind}
{Shlosman} I.,  {Vitello} P.,  1993, \mn@doi [\apj] {10.1086/172670}, \href {https://ui.adsabs.harvard.edu/abs/1993ApJ...409..372S} {409, 372}

\bibitem[\protect\citeauthoryear{{Shlosman}, {Vitello}  \& {Mauche}}{{Shlosman} et~al.}{1996}]{shl96v347pup}
{Shlosman} I.,  {Vitello} P.,   {Mauche} C.~W.,  1996, \mn@doi [\apj] {10.1086/177066}, \href {https://ui.adsabs.harvard.edu/abs/1996ApJ...461..377S} {461, 377}

\bibitem[\protect\citeauthoryear{{Sim}, {Drew}  \& {Long}}{{Sim} et~al.}{2005}]{sim05ysospec}
{Sim} S.~A.,  {Drew} J.~E.,   {Long} K.~S.,  2005, \mn@doi [\mnras] {10.1111/j.1365-2966.2005.09472.x}, \href {https://ui.adsabs.harvard.edu/abs/2005MNRAS.363..615S} {363, 615}

\bibitem[\protect\citeauthoryear{{Sion}}{{Sion}}{1995}]{sio95CVWDheating}
{Sion} E.~M.,  1995, \mn@doi [\apj] {10.1086/175129}, \href {https://ui.adsabs.harvard.edu/abs/1995ApJ...438..876S} {438, 876}

\bibitem[\protect\citeauthoryear{{Stehle}, {King}  \& {Rudge}}{{Stehle} et~al.}{2001}]{ste01zcamstandstill}
{Stehle} R.,  {King} A.,   {Rudge} C.,  2001, \mn@doi [\mnras] {10.1046/j.1365-8711.2001.04223.x}, \href {https://ui.adsabs.harvard.edu/abs/2001MNRAS.323..584S} {323, 584}

\bibitem[\protect\citeauthoryear{{Szkody} et~al.,}{{Szkody} et~al.}{2013}]{szk13V455And}
{Szkody} P.,  et~al., 2013, \mn@doi [\apj] {10.1088/0004-637X/775/1/66}, \href {https://ui.adsabs.harvard.edu/abs/2013ApJ...775...66S} {775, 66}

\bibitem[\protect\citeauthoryear{{Tampo} et~al.,}{{Tampo} et~al.}{2021}]{tam21seimeiCVspec}
{Tampo} Y.,  et~al., 2021, \mn@doi [\pasj] {10.1093/pasj/psab036}, \href {https://ui.adsabs.harvard.edu/abs/2021PASJ...73..753T} {73, 753}

\bibitem[\protect\citeauthoryear{{Tampo} et~al.,}{{Tampo} et~al.}{2022}]{tam22v455andspec}
{Tampo} Y.,  et~al., 2022, \mn@doi [\pasj] {10.1093/pasj/psac007}, \href {https://ui.adsabs.harvard.edu/abs/2022PASJ...74..460T} {74, 460}

\bibitem[\protect\citeauthoryear{{Thorstensen}, {Ringwald}, {Wade}, {Schmidt}  \& {Norsworthy}}{{Thorstensen} et~al.}{1991a}]{tho91pxand}
{Thorstensen} J.~R.,  {Ringwald} F.~A.,  {Wade} R.~A.,  {Schmidt} G.~D.,   {Norsworthy} J.~E.,  1991a, \mn@doi [\aj] {10.1086/115874}, \href {https://ui.adsabs.harvard.edu/abs/1991AJ....102..272T} {102, 272}

\bibitem[\protect\citeauthoryear{{Thorstensen}, {Davis}  \& {Ringwald}}{{Thorstensen} et~al.}{1991b}]{tho91bhlyn}
{Thorstensen} J.~R.,  {Davis} M.~K.,   {Ringwald} F.~A.,  1991b, \mn@doi [\aj] {10.1086/115902}, \href {https://ui.adsabs.harvard.edu/abs/1991AJ....102..683T} {102, 683}

\bibitem[\protect\citeauthoryear{{Tomaru}, {Done}, {Ohsuga}, {Nomura}  \& {Takahashi}}{{Tomaru} et~al.}{2019}]{tom19thermalradwind}
{Tomaru} R.,  {Done} C.,  {Ohsuga} K.,  {Nomura} M.,   {Takahashi} T.,  2019, \mn@doi [\mnras] {10.1093/mnras/stz2738}, \href {https://ui.adsabs.harvard.edu/abs/2019MNRAS.490.3098T} {490, 3098}

\bibitem[\protect\citeauthoryear{{Tovmassian}, {G{\"a}nsicke}, {Echevarria}, {Zharikov}  \& {Ramirez}}{{Tovmassian} et~al.}{2022}]{tov22v455andspec}
{Tovmassian} G.,  {G{\"a}nsicke} B.~T.,  {Echevarria} J.,  {Zharikov} S.,   {Ramirez} A.,  2022, \mn@doi [\apj] {10.3847/1538-4357/ac930a}, \href {https://ui.adsabs.harvard.edu/abs/2022ApJ...939...14T} {939, 14}

\bibitem[\protect\citeauthoryear{{Townsley} \& {G{\"a}nsicke}}{{Townsley} \& {G{\"a}nsicke}}{2009}]{tow09CVWDtemp}
{Townsley} D.~M.,  {G{\"a}nsicke} B.~T.,  2009, \mn@doi [\apj] {10.1088/0004-637X/693/1/1007}, \href {https://ui.adsabs.harvard.edu/abs/2009ApJ...693.1007T} {693, 1007}

\bibitem[\protect\citeauthoryear{{Vitello} \& {Shlosman}}{{Vitello} \& {Shlosman}}{1993}]{vit93cvwind}
{Vitello} P.,  {Shlosman} I.,  1993, \mn@doi [\apj] {10.1086/172799}, \href {https://ui.adsabs.harvard.edu/abs/1993ApJ...410..815V} {410, 815}

\bibitem[\protect\citeauthoryear{{Warner}}{{Warner}}{1995}]{war95book}
{Warner} B.,  1995, {Cataclysmic variable stars}.
Cambridge: Cambridge University Press

\bibitem[\protect\citeauthoryear{{Williams}}{{Williams}}{1989}]{wil89CVeclipse}
{Williams} R.~E.,  1989, \mn@doi [\aj] {10.1086/115115}, \href {https://ui.adsabs.harvard.edu/abs/1989AJ.....97.1752W} {97, 1752}

\bibitem[\protect\citeauthoryear{{Woods}, {Drew}  \& {Verbunt}}{{Woods} et~al.}{1990}]{woo90suumarxandbzcamUV}
{Woods} A.~J.,  {Drew} J.~E.,   {Verbunt} F.,  1990, \mnras, \href {https://ui.adsabs.harvard.edu/abs/1990MNRAS.245..323W} {245, 323}

\bibitem[\protect\citeauthoryear{{Woods}, {Klein}, {Castor}, {McKee}  \& {Bell}}{{Woods} et~al.}{1996}]{woo96thermalwind}
{Woods} D.~T.,  {Klein} R.~I.,  {Castor} J.~I.,  {McKee} C.~F.,   {Bell} J.~B.,  1996, \mn@doi [\apj] {10.1086/177101}, \href {https://ui.adsabs.harvard.edu/abs/1996ApJ...461..767W} {461, 767}

\bibitem[\protect\citeauthoryear{{Zorotovic}, {Schreiber}  \& {G{\"a}nsicke}}{{Zorotovic} et~al.}{2011}]{zor11SDSSCVWDmass}
{Zorotovic} M.,  {Schreiber} M.~R.,   {G{\"a}nsicke} B.~T.,  2011, \mn@doi [\aap] {10.1051/0004-6361/201116626}, \href {https://ui.adsabs.harvard.edu/abs/2011A&A...536A..42Z} {536, A42}

\bibitem[\protect\citeauthoryear{{{\v{S}}imon}}{{{\v{S}}imon}}{2003}]{sim03ssxsoptical}
{{\v{S}}imon} V.,  2003, \mn@doi [\aap] {10.1051/0004-6361:20030655}, \href {https://ui.adsabs.harvard.edu/abs/2003A&A...406..613S} {406, 613}

\makeatother
\end{thebibliography}








\bsp	
\label{lastpage}
\end{document}